\documentclass[epj]{svjour}
\usepackage{graphics}

\usepackage{mathptmx}      
\usepackage{flushend}

\usepackage{subfig}

\usepackage{amsfonts}
\usepackage{amssymb}

\begin{document}
\title{Frequentist limit setting in effective field theories}
\author{K. D. Gregersen\inst{1} \and J. B. Hansen\inst{2}
}                     
%
%
\institute{Department of Physics and Astronomy, University College London, Gower Street, London, WC1E 6BT, United Kingdom 
  \and Discovery Centre, Niels Bohr Institute, Faculty of Science, University of Copenhagen, Blegdamsvej 17, 2100 Copenhagen, Denmark 
}
\date{Received: date / Revised version: date}
%
\abstract{
The original frequentist approach for computing confidence intervals involves the construction of the confidence belt which provides a mapping of the observation in data into a subset of values for the parameter. 
There are different prescriptions for constructing the confidence belt, here we use the one provided by Feldman and Cousins.
Alternative methods based on the frequentist idea exist, including the delta likelihood method, the $CL_s$ method and a method here referred to as the $p$-value method, which have all been commonly used in high energy experiments. The purpose of this article is to draw attention to a series of potential problems when applying these alternative methods to the important case where the predicted signal depends quadratically on the parameter of interest, a situation which is common in high energy physics as it covers scenarios encountered in effective theories. These include anomalous Higgs couplings and anomalous trilinear and quartic gauge couplings. It is found that the alternative methods, contrary to the original method using the confidence belt, 
encode the goodness-of-fit into the confidence intervals and potentially over-constrain the parameter. 
%
} 
\maketitle
\section{Introduction}
\label{sec:intro}
The phenomenological description of Beyond the Standard \linebreak Model (BSM) physics in model independent searches is typically done in the framework of effective Lagrangians. The basic assumption is that there exists new physics with degrees of freedom so heavy that they cannot be produced directly at present colliders such as the Large Hadron Collider (LHC). The only observable effect is the modification of existing interactions or the introduction of new interactions between the Standard Model (SM) particles. These interactions are introduced by adding new terms with associated couplings to the SM Lagrangian; examples include anomalous Higgs couplings \cite{corbett} and anomalous trilinear \cite{hagiwara} and quartic \cite{quartic} gauge couplings. The new terms in the Lagrangian are typically non-renormalis-able which makes the differential cross section increase as function of energy and eventually violate S-matrix unitarity \cite{uni1,uni2,uni3}.

Since the new couplings enter linearly in the Lagrangian, the differential cross section depends quadratically on the couplings through the amplitude squared. The parabolic behaviour of the differential cross section implies the existence of a lower bound on the predicted signal. For the cases studied at the LHC, such as anomalous Higgs couplings and anomalous trilinear and quartic gauge couplings, this bound is typically located close to or at the SM expectation. Consequently, experimental outcomes which show distinct downward fluctuations with respect to the SM expectation are not described by the model. 

The inadequacy of the model to describe all experimental outcomes does not indicate that the model is wrong, but rather that it is sensitive to statistical fluctuations in a finite data sample. It should also be emphasised that the parameter of the model is unbound in both a physical and a mathematical sense. The problem is therefore \textit{not} concerned with a parameter boundary.

The study presented here is related to previous work in the literature considering problems with quadratic parameter dependence, where the parameter \textit{is} bound, e.g. the well-known problem of measuring neutrino masses \cite{feldmancousins}. However, in our case, the parameter dependence is more complex due to the presence of a linear term. For this reason, the standard approach of restating the bound on the differential cross section as a bound on the parameter squared cannot be employed. 

We review a number of statistical methods currently used for estimating couplings in effective field theories highlighting differences in the resulting confidence intervals.
A comprehensive treatment of confidence intervals is given in \cite{yellow}, including the case of couplings in effective field theories \cite{kersevan} where the approach by Feldman and Cousins \cite{feldmancousins} is presented, albeit, in a form which does not ensure proper confidence intervals. 

This article is organised as follows: Section \ref{sec:signal} describes the theoretical bound on the predicted signal coming from the quadratic parameter dependence for BSM contributions in effective theories. Section \ref{sec:methods} presents the most commonly used frequentist methods for determining confidence intervals. Section \ref{sec:baur} introduces a set of distributions called the \emph{Baur set} which systematically probes different regions in the observable including those not described by the model due to the bound. The Baur set is used in section \ref{sec:comparison} for comparing the statistical methods for the special case where the interference between the SM and BSM terms is zero. In section \ref{sec:interference}, results are shown for the general case with non-zero interference. Section \ref{sec:conclusion} gives the conclusion.

\section{Theoretical bounds on the predicted signal}
\label{sec:signal}
In effective theories where the SM Lagrangian is extended with an extra interaction term and a corresponding coupling strength parameter, $\theta$, the differential cross section, ${\rm d}\sigma/{\rm d}x$, for a given observable, $x$, depends quadratically on the parameter through the amplitude squared,
\begin{eqnarray}\label{eqn:ampl_squared}
  \frac{{\rm d}\sigma}{{\rm d}x}(\theta) & \propto & |A_\mathrm{SM}(x) + A_\mathrm{BSM}(x)\cdot\theta|^2,
\end{eqnarray}
where $A_\mathrm{SM}(x)$ and $A_\mathrm{BSM}(x)$ denote the SM and BSM complex amplitudes, respectively, and the dependence on $\theta$ has been factored out from the BSM amplitude.

More explicitly, this means that the differential cross section can be written on the quadratic form
\begin{eqnarray}\label{eqn:signal}
  \frac{{\rm d}\sigma}{{\rm d}x}(\theta) &=& a_0(x) + a_1(x)\cdot\theta + a_2(x)\cdot\theta^2,
\end{eqnarray}
where $a_i(x)$ are real numbers depending on $x$ and integrated over all remaining phase space dependencies.

The first term, $a_0(x)$, denotes the point of expansion which is equivalent to the SM expectation. The coefficient in the linear term, $a_1(x)$, represents the interference between the SM and the BSM terms in the Lagrangian. The coefficient in the quadratic term, $a_2(x)$, solely contains the contribution from the BSM term in the Lagrangian. 

The parabolic behaviour in equation \ref{eqn:signal} implies a bound on the differential cross section, the observable effects of which depend on the signs and the relative sizes of $a_0(x)$, $a_1(x)$ and $a_2(x)$. 

The sign of $a_2(x)$ determines whether the bound is a maximum or a minimum. In effective field theories, the non-renor-malisability of the BSM contribution usually renders $a_2(x)$ positive such that the bound introduced is a lower bound. For the following discussion, we assume that this is the case, but also note that an upper bound would give rise to the same conclusions.

If $a_1(x)$ is relatively large\footnote{The allowed range of $a_1(x)$ will be discussed in section \ref{sec:interference}.} compared to $a_2(x)$, the extremum in ${\rm d}\sigma/{\rm d}x$ is shifted significantly away from $\theta=0$ and the signal prediction behaves pseudo linearly for small $|\theta|$. In this case, the model is able to describe experimental outcomes with event yields below the SM expectation, which means that the bound on the differential cross section has a small effect as long as the observation is not too far from the SM expectation relative to the sensitive. However, the linear term is often very small\footnote{In fact, the linear term is completely absent if the BSM terms in the Lagrangian are CP violating.} compared to the quadratic term and hence the bound is close to $\theta=0$. Consequently, even relatively small fluctuations away from the SM expectation can result in an observation which is not described by the model. A scenario of this kind is the main focus of this article. 

The quadratic parameter dependence can be generalised to an expansion of any power larger than or equal to two, corresponding to a higher order operator expansion,
\begin{eqnarray}\label{eqn:signal2}
  \frac{{\rm d}\sigma}{{\rm d}x}(\theta) &=& \sum_{i=0}^n a_i(x)\cdot\theta^i \quad , \quad n\geq 2.
\end{eqnarray}
The exact number of non-zero coefficients $a_i(x)$ is not important to the arguments presented here as long as the highest power in the expansion is even. If the highest power is odd, many of the considerations presented here are still important depending on the specific physics model. In fact, since the important feature is that the model is unable to describe all possible experimental outcomes, the results presented here are not limited to a power law expansion, but are relevant for any function of $\theta$ for which this is the case.

\section{Determination of confidence intervals}
\label{sec:methods}
In high energy physics, the most commonly used methods for computing confidence intervals are the confidence belt, the delta likelihood method, the $CL_s$ method, and a method here referred to as the $p$-value method. This section gives brief descriptions of these methods with emphasis on the specific properties which are special for scenarios where the signal prediction depends quadratically on the parameter of interest. 

First, it is useful to recall the definition of a confidence interval. A confidence interval is an interval estimate of a model parameter which contains the unknown true value of the parameter with a probability given by the \emph{confidence level}. This means that in the limit where the experiment is repeated an infinite number of times and the confidence interval is recomputed every time, the probability that any of these confidence intervals contains the true value of the parameter is equal to the confidence level. 

This leads to the important concept of \textit{coverage probability}. Coverage probability is the proportion of the time that the confidence interval contains the true value of the parameter. Thus, it can be regarded as the actual confidence level of the computed interval. Ideally, the coverage probability is equal to the confidence level. If the coverage probability is smaller than the confidence level, the confidence interval is termed \emph{permissive}, while it is termed \emph{conservative} if the coverage probability is greater than the confidence level.

For illustrative purposes, a binned observable $x$ is introduced using 10 bins in the range $[0;1]$. The number of measurements, also referred to as the number of \textit{events}, in each bin of the observable is governed by Poisson statistics. The likelihood function is defined as the product of the probabilities for the individual Poisson processes, i.e. 
\begin{eqnarray}\label{eqn:likelihood}
  \mathcal{L}(\theta) \equiv \displaystyle{\prod_{i=1}^{N}}\frac{\mu_i^{n_i}(\theta)}{n_i!}e^{-\mu_i(\theta)},
\end{eqnarray}
where the expected number of events in the $i^{\textrm{\tiny th}}$ bin is given by $\mu_i(\theta)$ which depends quadratically on $\theta$, i.e. 
\begin{eqnarray}\label{eqn:signal2}
  \mu_i(\theta) &=& a_{0,i} + a_{1,i}\cdot\theta + a_{2,i}\cdot\theta^2.  
\end{eqnarray}

The quantities $n_i$ in equation \ref{eqn:likelihood}  correspond to the number of observed events in the $i^{\textrm{\tiny th}}$ bin of the observable and the set $\{n_i\}$ is referred to as the \textit{observation}. For pseudo data, $\{n_i\}$ are integers, but when using the predicted signal for a given value of $\theta$ as the event count, e.g. when estimating the confidence interval for the SM expectation, $\{n_i\}$ are treated as real numbers\footnote{In this case, the factorial in the Poisson probability in equation \ref{eqn:likelihood} is substituted with the Gamma function.}. 

It should be noted that in equation \ref{eqn:likelihood}, we have adopted the standard approach where the likelihood is considered a function of the parameter, hence suppressing the dependence on $\{n_i\}$ in the notation, i.e.
\begin{eqnarray}
  \mathcal{L}(\theta) \equiv \mathcal{L}(\{n_i\}|\theta). 
\end{eqnarray}

\begin{figure}
  \centering
  \resizebox{0.88\columnwidth}{!}{\includegraphics{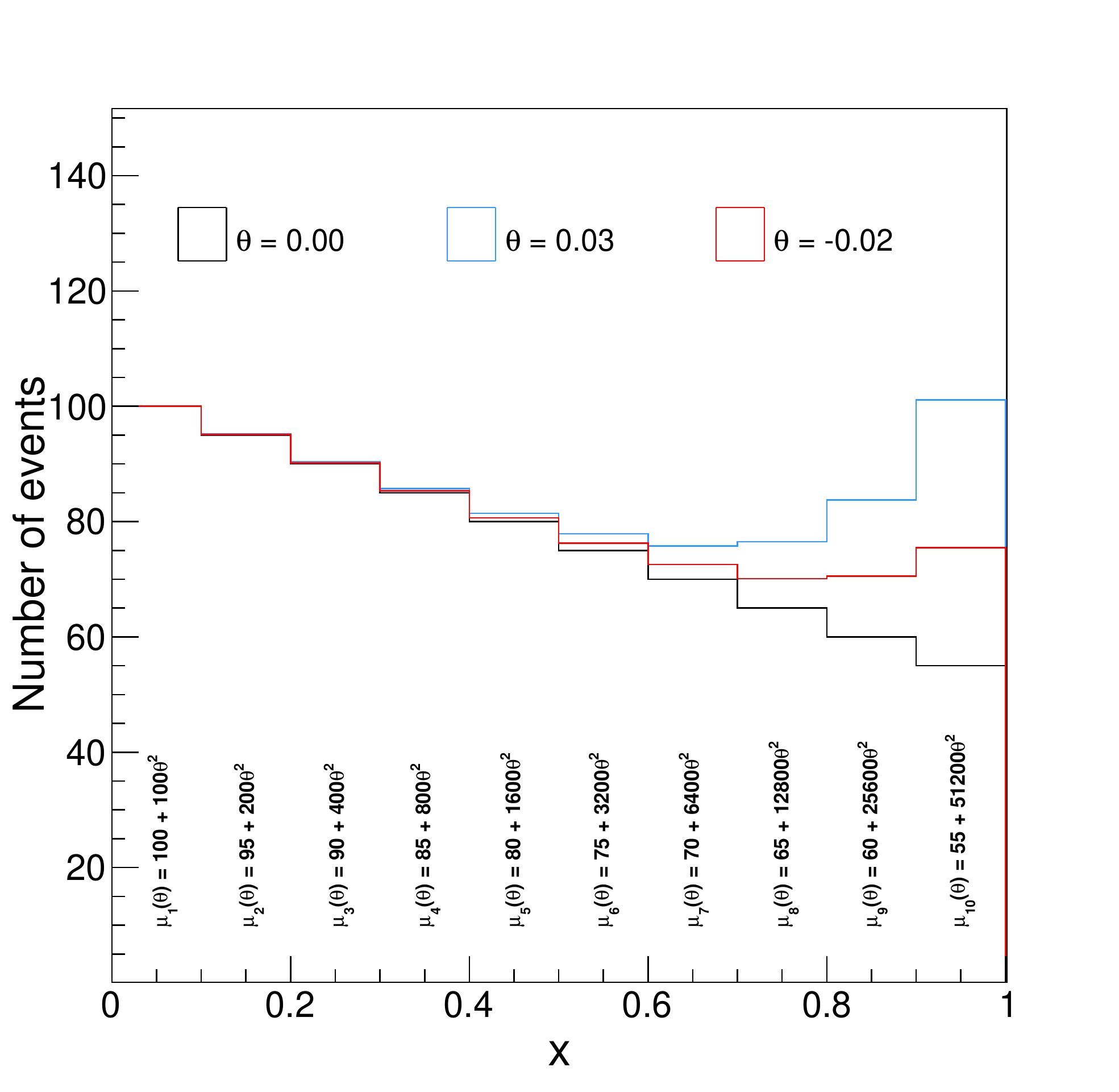}}
  \caption{Distributions of the observable, $x$, for 3 values of the parameter: $\theta = 0$ (black), $\theta = 0.03$ (blue) and $\theta = -0.02$ (red). The signal parameterisation has $a_1 = 0$ for all bins in the observable. For convenience, the values of $a_0$ and $a_2$ are given below the graphs for each bin, respectively. }
  \vspace{0.1cm}
  \label{fig:observable}  
\end{figure}

Initially, the interference term in the model is set to zero for all values of the observable, i.e. $a_{1} = 0$ for all bins\footnote{The effects of non-zero interference, $a_1(x) \neq 0$, are addressed in section \ref{sec:interference}.}. The value of $a_{2}$ is increasing across the interval in $x$ such that the sensitivity of $x$ to the parameter $\theta$ grows monotonically with $x$ \footnote{This choice is arbitrary. Here we have chosen values that give a behaviour similar to what is encountered in LHC and Tevatron experiments.}. 
Figure \ref{fig:observable} shows the SM expectation (black) and the distributions for $\theta = 0.03$ (blue) and $\theta = -0.02$ (red) using this parameterisation. The values of $a_0$ and $a_2$ in each bin is given in the plot. The sizes of all data samples are large enough to avoid dealing with features related to low event counts in the individual bins of the observable. 

All fits are performed using the minimisation routine \linebreak{\verb MINUIT } \cite{Minuit} via its implementation in {\verb ROOT } \cite{ROOT} and the function which is minimised is twice the negative logarithm of the likelihood ratio, defined as
\begin{eqnarray}\label{eqn:-2lnq}
  -2\ln q(\theta) &\equiv& -2[\ln \mathcal{L}(\theta) - \ln \mathcal{L}(\hat{\theta})],
\end{eqnarray}
where $q$ denotes the likelihood ratio and $\hat{\theta}$ is the maximum likelihood estimator for $\theta$.

The next sub-sections describe the four statistical methods and will as illustration use the SM expectation as the observation. 

\subsection{Delta likelihood method}
\label{sec:deltaLLR}

\begin{figure}
  \centering
  \vspace{-0.3cm}
  \resizebox{0.95\columnwidth}{!}{\includegraphics{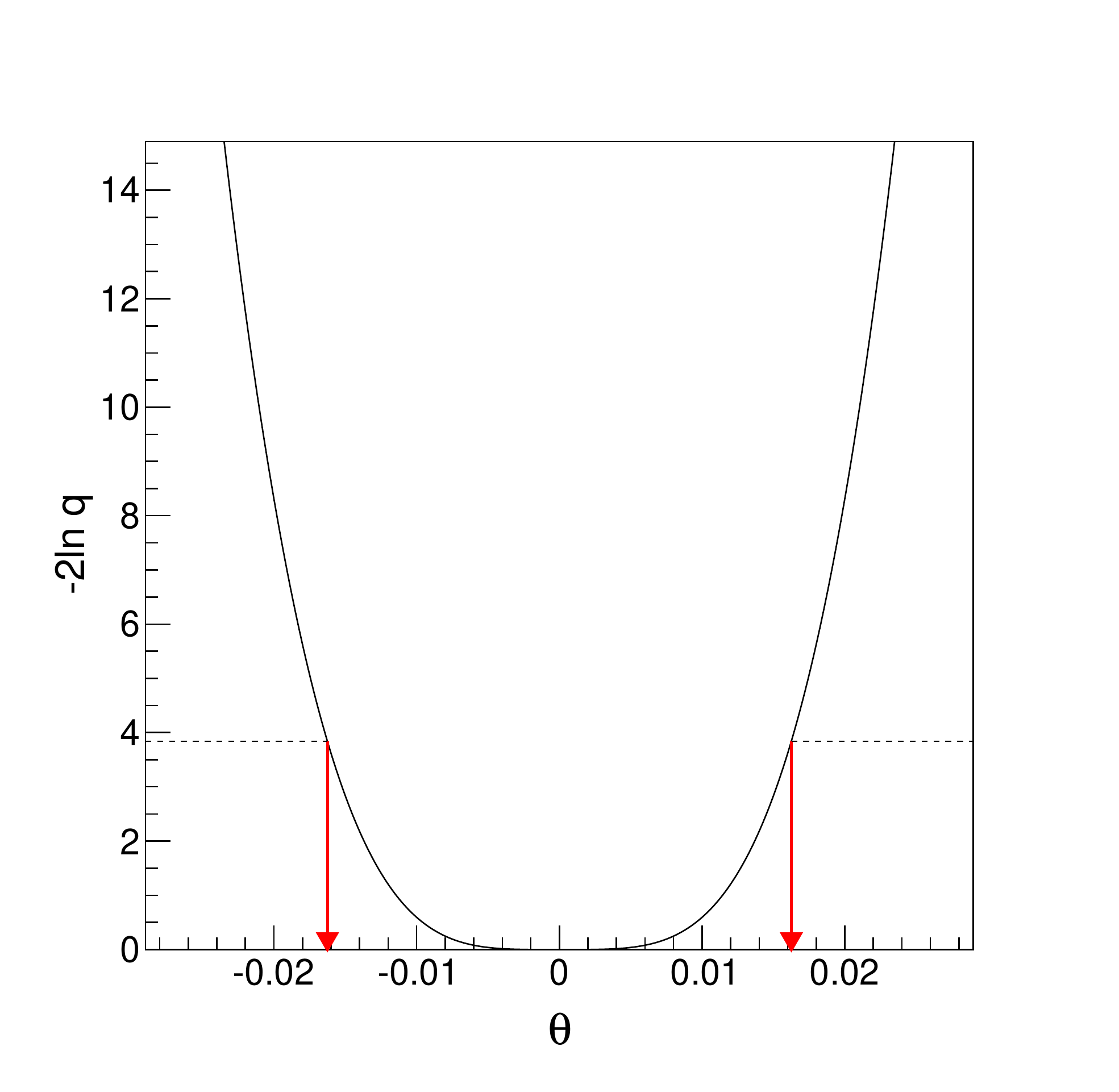}}
  \vspace{-0.3cm}
  \caption{The black curve shows $-2\ln q$ as function of $\theta$ when the SM expectation is used as the observation. The intersections between the horizontal line at 3.84 and the curve gives the delta likelihood ratio interval at 95\% CL as indicated by the vertical arrows.} 
  \label{fig:profile_casestudy} 
\end{figure}

Traditionally, the delta likelihood method has been used for reporting confidence intervals, e.g. \cite{D0}, \cite{lep}. It is the simplest and fastest method for computing confidence intervals among the approaches described here, since it does not require large amounts of simulated data. 

The confidence interval is estimated by considering the variation of the likelihood function near its maximum. It is given by the interval $[\theta_{\rm low},\theta_{\rm high}]$ for which $\theta$ satisfies 
\begin{eqnarray}
  -2\ln q(\theta) &<& -2\ln q_{\rm CL},
\end{eqnarray}
where $-2\ln q_{\rm CL}$ is a constant computed from the chi-square distribution with one free parameter. For a confidence interval at 95\% CL, this is given by $-2\ln q_{\rm 95\%} = 3.84$. 

Figure~\ref{fig:profile_casestudy} shows $-2\ln q$ when the SM expectation is used as the observation. The dashed horizontal line indicates the 95\% CL and the vertical arrows give the end-points of the corresponding confidence interval.

\subsection{Confidence belt}

\begin{figure}
  \centering
  \vspace{-0.5cm}
  \resizebox{0.945\columnwidth}{!}{\includegraphics{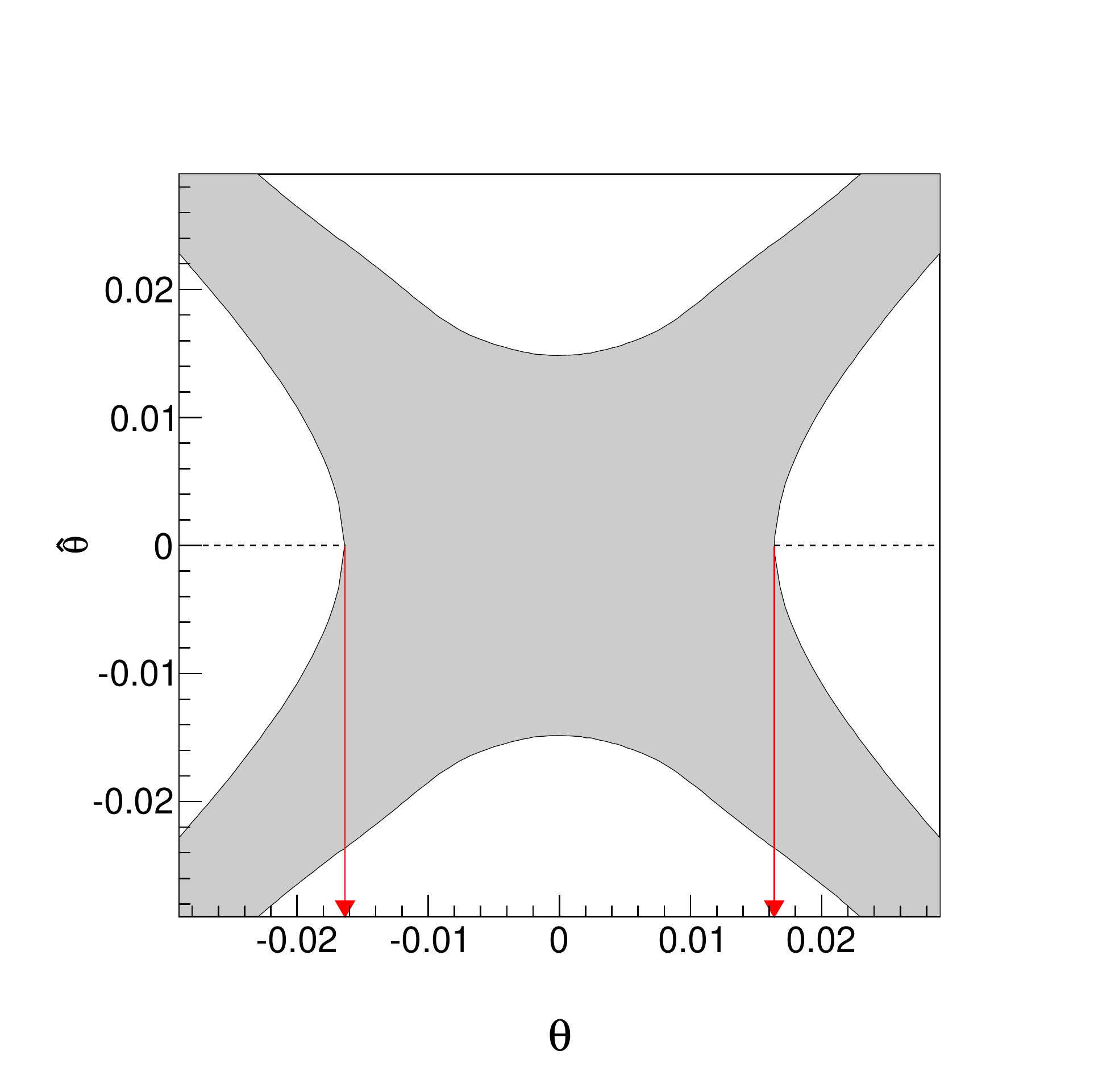}}
  \caption{The contour plot shows the Neyman construction at 95\% CL with Feldman-Cousins ordering. The confidence interval is given by the intersections with the dashed horizontal line at $\hat{\theta}_{\rm obs}=0$ as indicated by the vertical arrows.}
  \label{fig:neyman_a} 
\end{figure}

The original frequentist confidence interval $[\theta_{\rm low},\theta_{\rm high}]$ for the parameter $\theta$ is computed by constructing the confidence belt. This is also called a Neyman construction as the general principle was first formulated by Jerzy Neyman in 1937 \cite{neyman}.

The confidence belt consists of the conjunction of intervals $[\hat{\theta}_{\rm low},\hat{\theta}_{\rm high}]$ which are determined for each value of $\theta$ by integrating the probability density function $P(\hat{\theta}|\theta)$ such that 
\begin{eqnarray}\label{eqn:neyman1}
  \int_{\hat{\theta}_{\rm low}}^{\hat{\theta}_{\rm high}}P(\hat{\theta}|\theta){\rm d}\hat{\theta} &=& \alpha,
\end{eqnarray}
where $\alpha$ denotes the confidence level. 

The belt has the property that as long as equation \ref{eqn:neyman1} is satisfied for all $\theta$, any orthogonal intersection with the confidence belt at a given $\hat{\theta}$ gives 
a set of intervals in $\theta$ with a coverage probability $\alpha$. Thus, the confidence interval is determined by the orthogonal intersection at the value of the maximum likelihood estimator for the observation, $\hat{\theta}_{\rm obs}$. 

While this procedure ensures coverage by construction, it still allows the freedom to choose which elements to be inside the interval given by equation \ref{eqn:neyman1}. The exact choice makes the interval unique and is known as the \emph{ordering principle}. Feldman and Cousins developed an ordering principle which usually is referred to as \emph{the unified approach}, \emph{likelihood ratio ordering} or \emph{Feldman-Cousins ordering} \cite{feldmancousins}. According to this principle, the interval is defined by including elements of probability ordered by their likelihood ratios such that higher ratios are given precedence over lower ratios for inclusion in the belt. 

The Feldman-Cousins ordering prescription is used here to to be able to make a direct comparison to the $p$-value method described in section \ref{sec:pvalue}. It should be noted that the original problems addressed by Feldman and Cousins are not present in our case. 

Figure~\ref{fig:neyman_a} shows the confidence belt at 95\% CL. It is constructed numerically with simulated data in the form of pseudo experiments drawn from the expected distribution for a suitable range in $\theta$. The distinct cross like shape of the confidence belt reflects the quadratic dependence on $\theta$ in the signal prediction which implies that $\theta$ is mapped to both same sign and opposite sign $\hat{\theta}$. When the SM expectation is used as the observation, the confidence interval is given by the intersections between the dashed horizontal line at $\hat{\theta}_{\rm obs}=0$ and the confidence belt as illustrated by the vertical arrows in figure~\ref{fig:neyman_a}.

\begin{figure}
  \centering
  \vspace{-0.5cm}
  \resizebox{0.945\columnwidth}{!}{\includegraphics{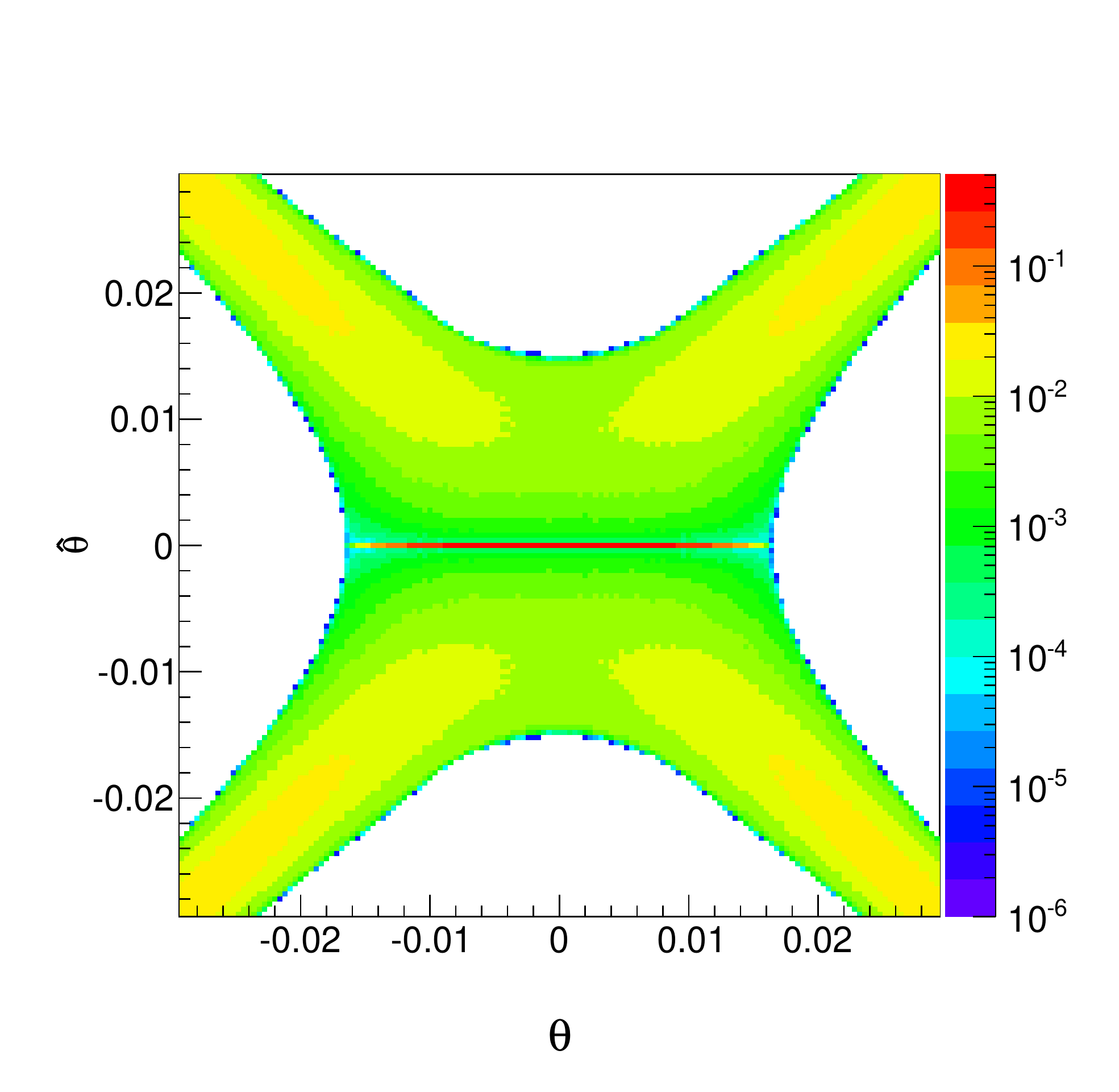}}
  \caption{The distribution of the 95\% highest ranking pseudo experiments using Feldman-Cousins ordering which defines the confidence belt shown in figure \ref{fig:neyman_a}.}
  \label{fig:neyman_b} 
\end{figure}

\begin{figure}
  \centering
  \vspace{-0.5cm}
  \resizebox{0.945\columnwidth}{!}{\includegraphics{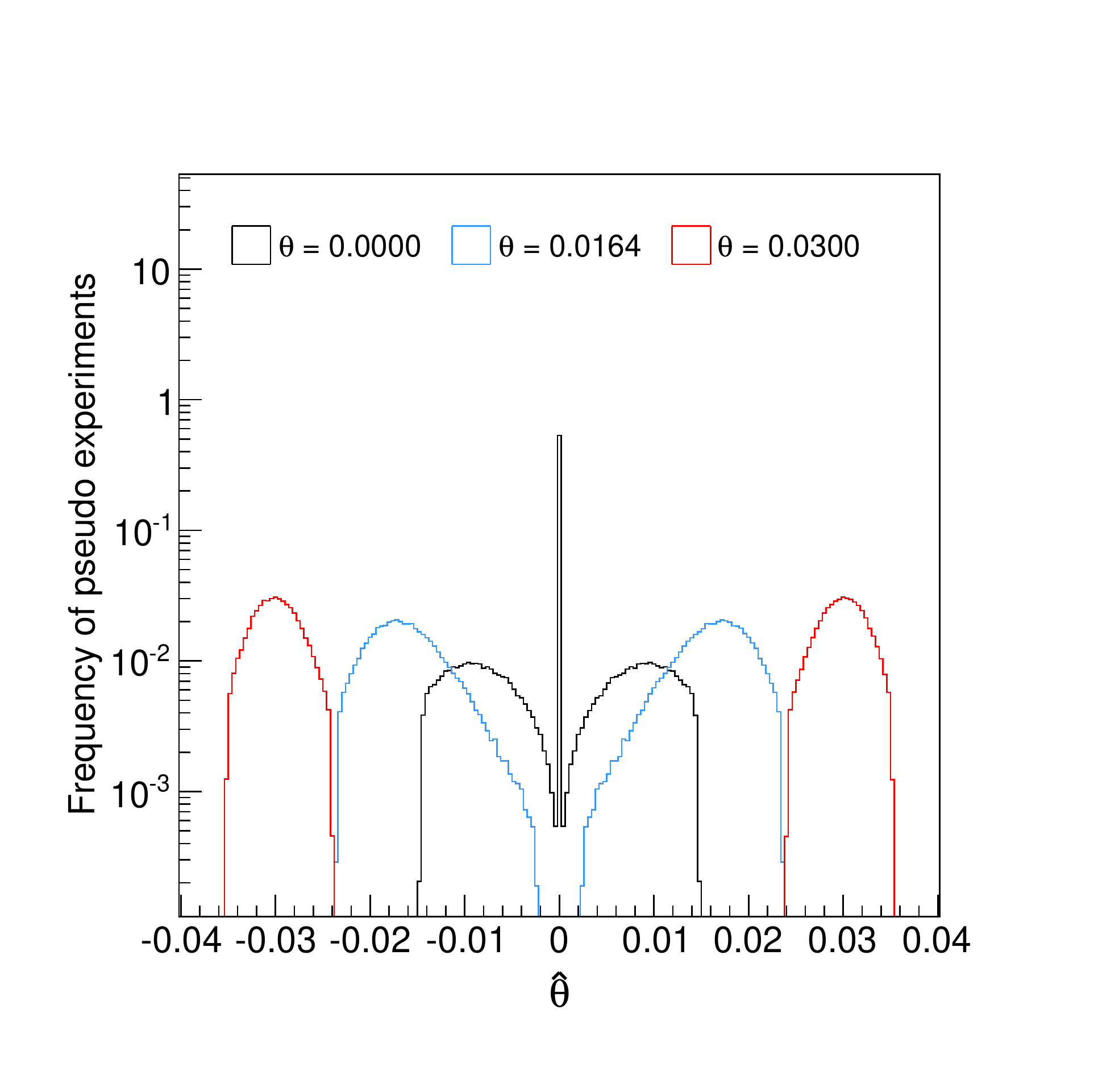}}
  \caption{Distributions of the 95\% highest ranking pseudo experiments for $\theta = 0$ (black), $\theta=0.0164$ (blue) and $\theta = 0.03$ (red), corresponding to three vertical projections at these values of $\theta$ of the contour plot shown in figure \ref{fig:neyman_b}.}
  \label{fig:neyman_c} 
\end{figure}

\begin{figure}
  \centering
  \vspace{-0.5cm}
  \resizebox{0.945\columnwidth}{!}{\includegraphics{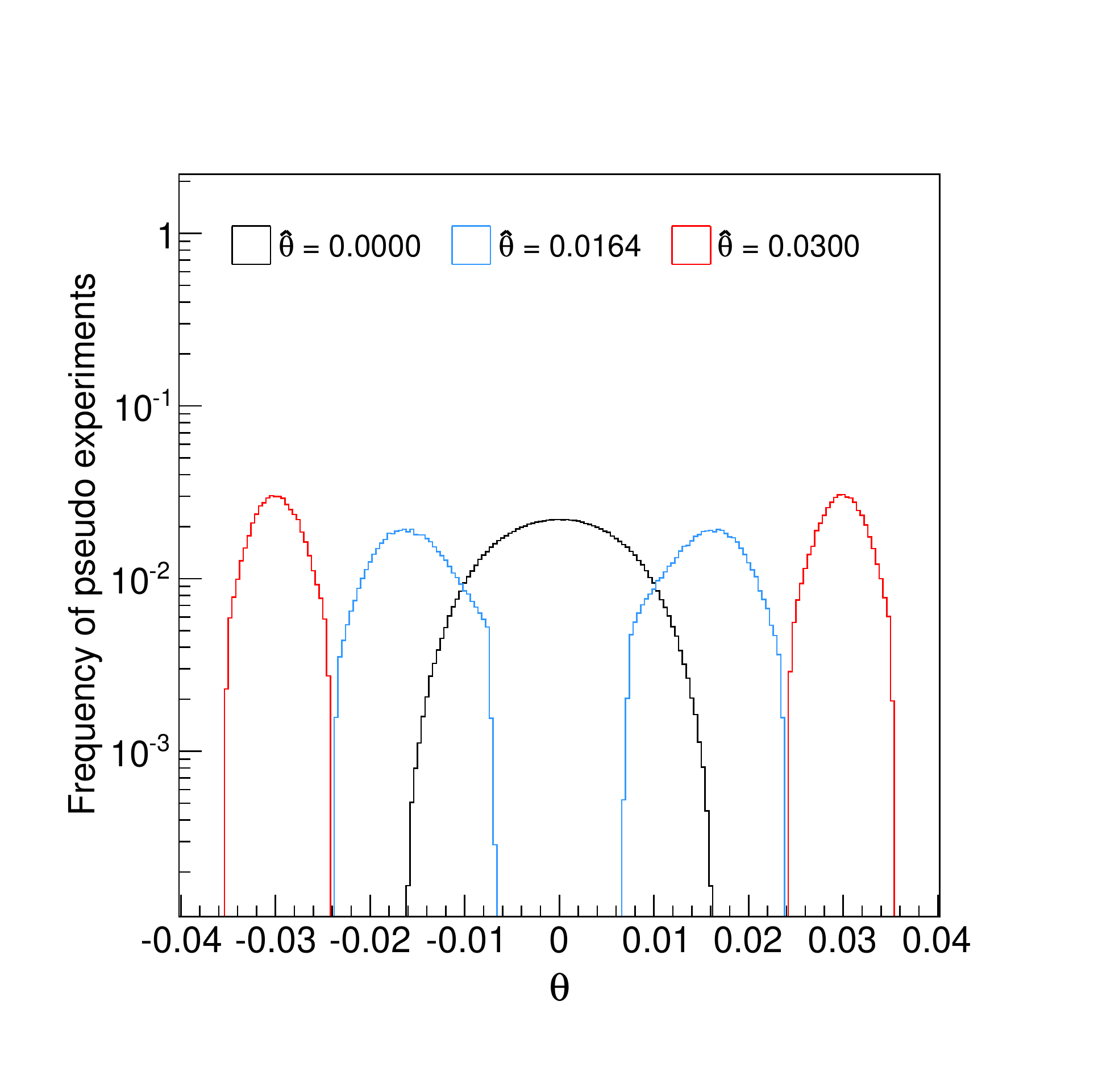}}
  \caption{Distributions of the 95\% highest ranking pseudo experiments for $\hat{\theta} = 0$ (black), $\hat{\theta}=0.0164$ (blue) and $\hat{\theta} = 0.03$ (red), corresponding to three horizontal projections at these values of $\hat{\theta}$ of the contour plot shown in figure \ref{fig:neyman_b}.}
  \label{fig:neyman_d} 
\end{figure}

As further illustration, figure \ref{fig:neyman_b} shows the two-dimensional distribution of the 95\% highest ranking pseudo experiments which define the confidence belt. It is seen that many pseudo experiments give $\hat{\theta} = 0$. This is also seen in figure \ref{fig:neyman_c} which shows three vertical projections of the two-dimensional distribution in figure \ref{fig:neyman_b}. The projection for $\theta=0$ (black histogram in figure \ref{fig:neyman_c}) shows that roughly half of the pseudo experiments give $\hat{\theta} = 0$. The reason for this sharp peak at $\hat{\theta} = 0$ is that the lower bound on the predicted signal is located at the SM, $\theta=0$, when the linear term is absent. For pseudo experiments generated around $\theta=0$, there is a high probability that a significant part of the bins in the observable have a downward fluctuation in the event yield wrt. the SM expectation and since these bins suggest that the best fit is $\hat{\theta} = 0$, the fit is pulled in this direction. Figure \ref{fig:neyman_d} shows three horisontal projections of the two-dimensional distribution in figure \ref{fig:neyman_b}. It is seen that these projections show no signs of the peak structure visible in figure \ref{fig:neyman_c} since the peak structure is purely horisontal. The horisontal projection at $\hat{\theta} = 0$ (black histogram in figure \ref{fig:neyman_d}) shows how the peak evolves as function of $\theta$ and that it is a smoothly rising distribution and symmetric around $\theta = 0$ as should be expected.

As mentioned in section \ref{sec:intro}, the Feldman-Cousins approach has been studied before in the context of couplings in effective field theories \cite{kersevan}. However, the specific implementation of the method is different from what is done here. In fact, in \cite{kersevan} two different but equivalent methods are employed for single- and multi-bin distributions, respectively. For the single-bin distribution, an observed event yield is used to derive a confidence interval on the predicted event yield via the Feldman-Cousins prescription. This interval is then translated into an interval on the parameter by solving the quadratic equation describing the relation between the two. The problem with this approach is that the translation does not preserve probability, as the mapping from the event yield to the parameter does not exist for all values of the event yield. 
In order to properly map between the observation in data and the true parameter, the observed event yield must first be stated in terms of the measured parameter before mapping into a subset of values for the true parameter, as we have done here. For a multi-bin distribution, the implementation in \cite{kersevan} reverts to the equivalent $p$-value method, described in the following.

\subsection{$p$-value method}
\label{sec:pvalue}

The $p$-value method is an alternative frequentist approach to the confidence belt. Traditionally, $p$-values are used for hypothesis testing and do not depend on the parameter. However, at the LHC it has been used to report confidence intervals in conjunction with parameter estimation, e.g. \cite{ATLAS_WZ}.

The idea is to determine the confidence interval by inverting a hypothesis test quantified by a $p$-value. This approach is completely equivalent to the confidence belt with likelihood ratio ordering when the signal prediction depends linearly on the parameter, which includes the important case of estimating the signal strength parameter in a resonance search. In this case, the confidence belt corresponds to the acceptance region of the hypothesis test\footnote{As will be demonstrated later, this relationship does not hold when the signal prediction depends quadratically on the parameter of interest.}.

The $p$-value is defined as
\begin{eqnarray}\label{eqn:pvaluedef}
  p(\theta) &\equiv& -2\int_{-2\ln q_{\rm obs}(\theta)}^\infty f(-2\ln q(\theta)){\rm d}\ln q(\theta),
\end{eqnarray}
where $f(-2\ln q(\theta))$ denotes the distribution of $-2\ln q$ for a given $\theta$.

The confidence interval is determined by the interval in $\theta$ for which the $p$-value is larger than $1-\alpha$, where $\alpha$ indicates the confidence level. 

The calculation of the $p$-value can be done numerically by performing pseudo experiments. In this case, it is given by the fraction of pseudo experiments for which the value of $-2\ln q$ is larger than it is for the observation, i.e.
\begin{eqnarray}\label{eqn:pvalue}
  p(\theta) &=& \frac{N_{-2\ln q_{\rm toy}(\theta)>-2\ln q_{\rm obs}(\theta)}}{N_{\rm total}},
\end{eqnarray}
where 
\begin{eqnarray}
  -2\ln q_{\rm toy}(\theta) &=& -2[\ln\mathcal{L}_{\rm toy}(\theta) - \ln\mathcal{L}_{\rm toy}(\hat{\theta})],
\end{eqnarray}
and
\begin{eqnarray}\label{eqn:likelihood_data}
  -2\ln q_{\rm obs}(\theta) &=& -2[\ln\mathcal{L}_{\rm obs}(\theta) - \ln\mathcal{L}_{\rm obs}(\hat{\theta})].
\end{eqnarray}
The likelihood functions for the observation and pseudo experiment are denoted $\mathcal{L}_{\rm obs}$ and $\mathcal{L}_{\rm toy}$, and $N_{-2\ln q_{\rm toy}(\theta)>-2\ln q_{\rm obs}(\theta)}$ is the number of pseudo experiments for which the value of $-2\ln q$ is larger than it is for the observation, while $N_{\rm total}$ is the total number of pseudo experiments performed for this value of $\theta$.

\begin{figure}
  \centering
  \resizebox{0.805\columnwidth}{!}{\includegraphics{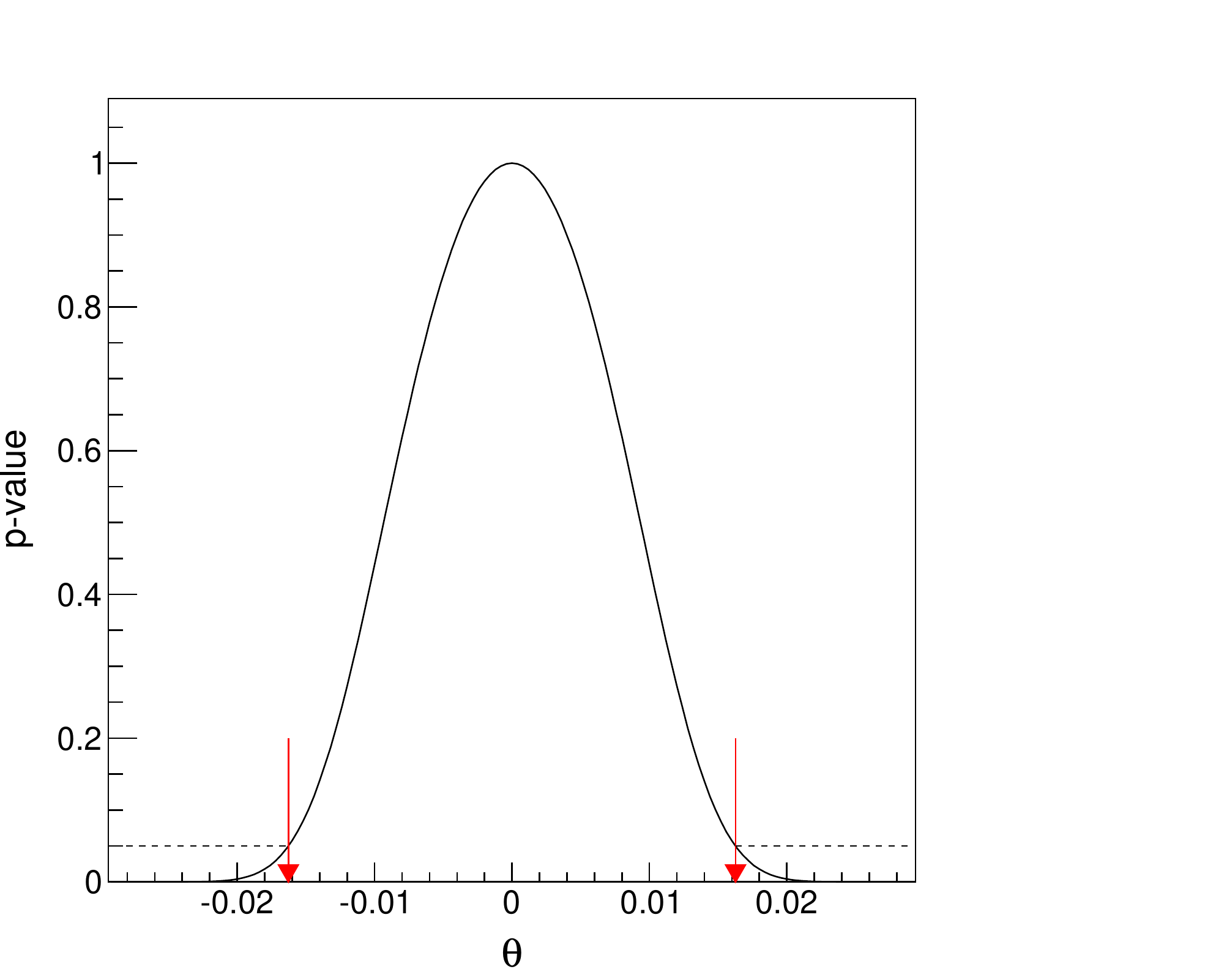} \hspace{-4cm}}
  \caption{The solid black curve shows the $p$-value when the SM expectation is used as the observation. The confidence interval is given by the values in $\theta$ for which the $p$-value is larger than $1-\alpha$ where $\alpha$ denotes the confidence level (shown as a dashed line for a 95\% CL). The end-points of the confidence interval at 95\% CL are indicated by the vertical arrows.}
  \label{fig:pvalue_casestudy_a} 
\end{figure}

In figure~\ref{fig:pvalue_casestudy_a}, the solid black curve shows the $p$-value as function of $\theta$ when the SM expectation is used as the observation. The vertical arrows indicate the values of $\theta$ for which the $p$-value is 5\% and thus determine the confidence interval at 95\% CL.

\begin{figure}
  \centering
  \resizebox{\columnwidth}{!}{\includegraphics{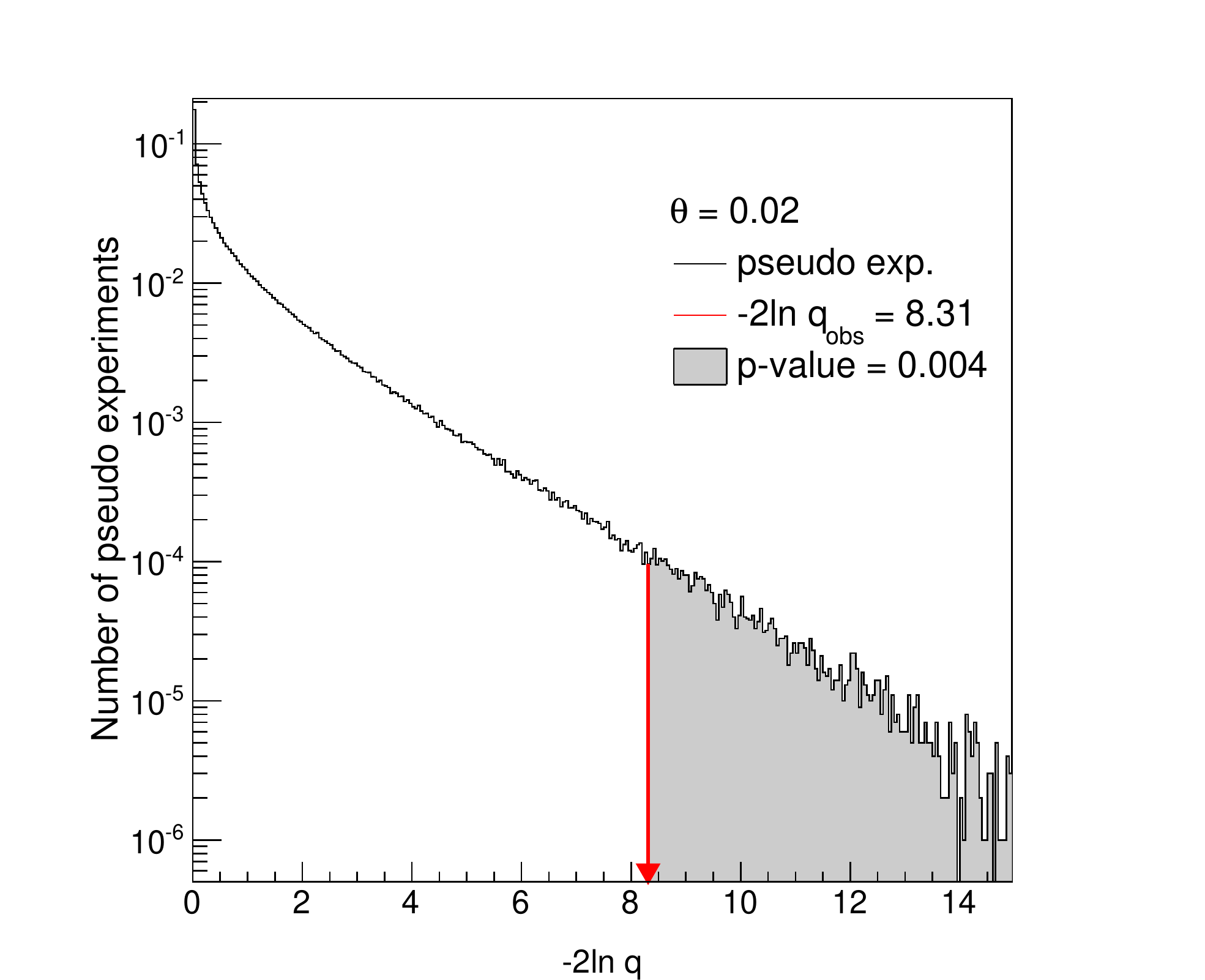}}
  \caption{The histogram shows the distribution of $-2\ln q$ for pseudo experiments produced for $\theta=0.02$. The vertical arrow indicates the value of $-2\ln q$ for the observation. The $p$-value is equal to the fraction of pseudo experiments which fall above this value, i.e. inside the grey area.}
  \label{fig:pvalue_casestudy_b} 
\end{figure}

In order to illustrate the $p$-value method in more detail, figure~\ref{fig:pvalue_casestudy_b} shows the distribution of $-2\ln q_{\rm toy}$ for a specific value of $\theta$. The vertical arrow indicates the corresponding value for the observation, $-2\ln q_{\rm obs}$, and the grey-shaded area represents the pseudo experiments which have $-2\ln q_{\rm toy}>-2\ln q_{\rm obs}$. The ratio between the number of pseudo experiments in the grey-shaded area and all pseudo experiments gives the $p$-value for this $\theta$. 

It should be noted that if the distribution of $-2\ln q_{\rm toy}$ follows a chi-square distribution with one free parameter for all $\theta$, the $p$-value and delta likelihood methods produce identical confidence intervals. This will be examined in more detail in section \ref{sec:comparison} where it will be shown that while the distribution of $-2\ln q$ indeed does follow a chi-square distribution for large values of $|\theta|$, the same is not true for values of $\theta$ around zero.

\subsection{$CL_s$ method}
\label{sec:cls}
The $CL_s$ method \cite{cls} was developed during the running of the Large Electron-Positron (LEP) collider and has been used both at LEP and at the LHC to report confidence intervals in resonance searches, e.g. \cite{lephiggs,ATLAShiggs,CMShiggs}, and parameters in effective theories, e.g. \cite{CMSatgc}. It is motivated by the attempt to provide more conservative confidence intervals in the case of a non-ob\-ser\-va\-tion where both the background-only, i.e. the SM, and the signal-plus-background hypotheses are disfavoured by the observation. For this reason, the $CL_s$ method is by construction not expected to give the correct frequentist coverage probability.

The $CL_s$ method proceeds by calculating $p$-values, as defined in equations \ref{eqn:pvaluedef}-\ref{eqn:pvalue}, for the background-only hypothesis, denoted $CL_b$, and the signal-plus-background hypothesis, denoted $CL_{s+b}(\theta)$. The quantity $CL_s(\theta)$ is then defined as the ratio between the $p$-values for the two hypotheses, 
\begin{eqnarray}\label{eqn:cls}
  CL_s(\theta) &\equiv& \frac{CL_{s+b}(\theta)}{CL_b}.
\end{eqnarray}

The confidence interval is determined by the values of $\theta$ for which $CL_s$ is larger than $1-\alpha$, where $\alpha$ denotes the confidence level.

When the SM expectation is used as the observation, the \linebreak $p$-value for the background-only hypothesis is exactly one, \linebreak $CL_b~=~1$. Consequently, the quantities denoted $p(\theta)$ and \linebreak $CL_s(\theta)$ in equations \ref{eqn:pvalue} and \ref{eqn:cls}, respectively, are the same and thus the $p$-value and $CL_s$ methods are identical. Section \ref{sec:comparison} investigates scenarios where this is not the case.

\section{The Baur Set}
\label{sec:baur}

As the problem under study arises for experimental outcomes not described by the model for any value of the parameter, we seek a procedure to define pseudo dataset in this region that reflects the parameter dependence in the allowed region. 

This may be achieved in many ways, here we choose a mapping related to the statistical sensitivity in the allowed region, in such a way that the migration between the two regions is exclusively tied to a single mapping parameter. The procedure works in any dimensionality and ensures well defined datasets for all values of the parameter. In short, this is achieved by scaling the SM distribution with the ratio between the SM and a distribution in the allowed region. The choice on parameter value for the distribution in the allowed region is done in terms of the statistical precision of the SM distribution. The resulting set of distributions are called the \textit{Baur set}\footnote{Named after late Ulrich Baur in recognition of his tremendous contribution to the field of diboson physics.}.

\begin{figure}
  \centering
  \vspace{-0.2cm}
  \resizebox{\columnwidth}{!}{\includegraphics{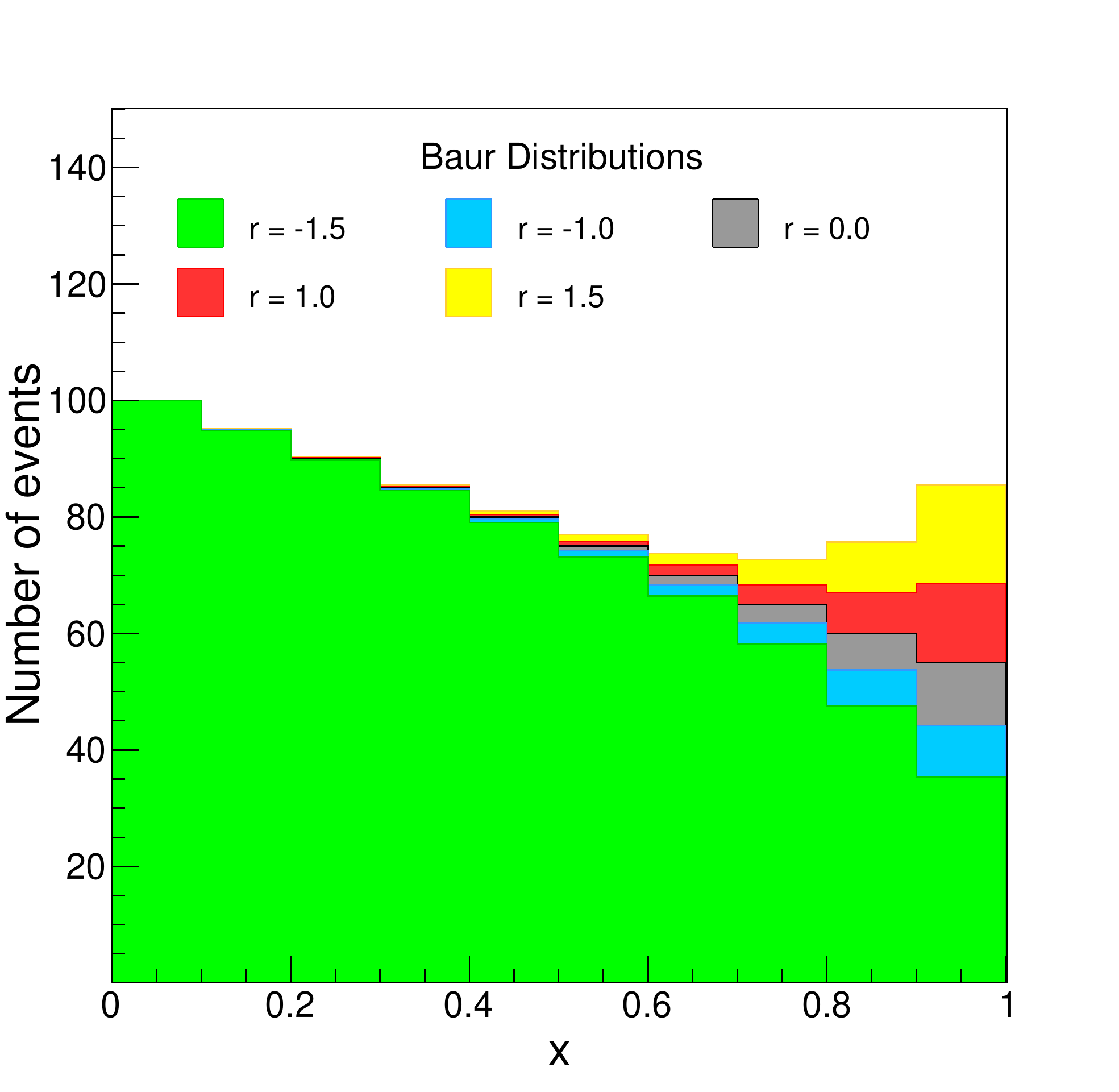}}
  \caption{Baur distributions for a subset of values of the Baur parameter, $r~\in~\{0, \pm 1, \pm 1.5\}$. The distributions are constructed with the value of $\sigma_\theta^{\rm ref}$ determined in section \ref{sec:deltaLLR}.}
    \label{fig:baurset}
\end{figure}

The Baur set consists of \textit{Baur distributions} which are uni-quely defined by their value of the \textit{Baur parameter}, $r$, which is a real number. The Baur distributions are constructed by first deriving the confidence interval $[\theta_{\rm low},\theta_{\rm high}]$ at 95\% CL as determined by the delta likelihood ratio using an observation at the SM expectation, and defining the quantity $\sigma_\theta^{\rm ref}$ as 
\begin{equation}\label{eqn:sigma_ref}
  \sigma_\theta^{\rm ref}~\equiv~(\theta_{\rm high}-\theta_{\rm low})/2.
\end{equation}
For a Gaussian likelihood function, this is simply two standard deviations\footnote{It should be noted that the choice of confidence level and statistical method for computing $\sigma_\theta^{\rm ref}$ is arbitrary, however, to keep it consistent with the choice of confidence level used in other sections, a 95\% CL is also used here, and the delta likelihood ratio is used in order to keep the definition as simple as possible from a computational point of view.}. 

Given this measure, the full Baur set is then defined as the infinite set of Baur distributions, $B(x;r)$, given by
\begin{equation} \label{eqn:baur}
  B(x;r) = 
  \left\{
    \begin{array}{rl}
      h(x;r\sigma_\theta^{\rm ref}) & \ ,\ \ r\geq 0\\
      h(x;0)\frac{H(x,{\rm d}x;0)}{H(x,{\rm d}x;r\sigma_\theta^{\rm ref})} & \ ,\ \ r<0
    \end{array}
  \right.
\end{equation}
where $h(x;\theta)$ is the distribution of the observable $x$ for a given $\theta$ and $H(x,{\rm d}x,\theta)$ is the cumulative distribution of $h(x;\theta)$ in the small interval d$x$, 
\begin{equation}
  H(x,{\rm d}x;\theta) = \int_x^{x+{\rm d}x}h(x';\theta){\rm d}x',
\end{equation}
subject to the requirement $H(x,{\rm d}x; \theta) > 0$.

For the binned observable used here, these definitions translate into:
\begin{itemize}
\item[$\bullet$] $\quad B(x;r)\rightarrow B_i(r)$\\
\item[$\bullet$] $\quad h(x;\theta)\rightarrow h_i(\theta)$\\
\item[$\bullet$] $\quad H(x,\textrm{d}x;\theta)\rightarrow h_i(\theta)\textrm{d}x_i$
\end{itemize}
where $i$ denotes the bin number, $h_i(\theta)$ is the event yield which is given by $\mu_i(\theta)$ in equation \ref{eqn:signal2}, and d$x_i$ is the bin width.

\begin{figure}
  \centering
  \resizebox{0.96\columnwidth}{!}{\includegraphics{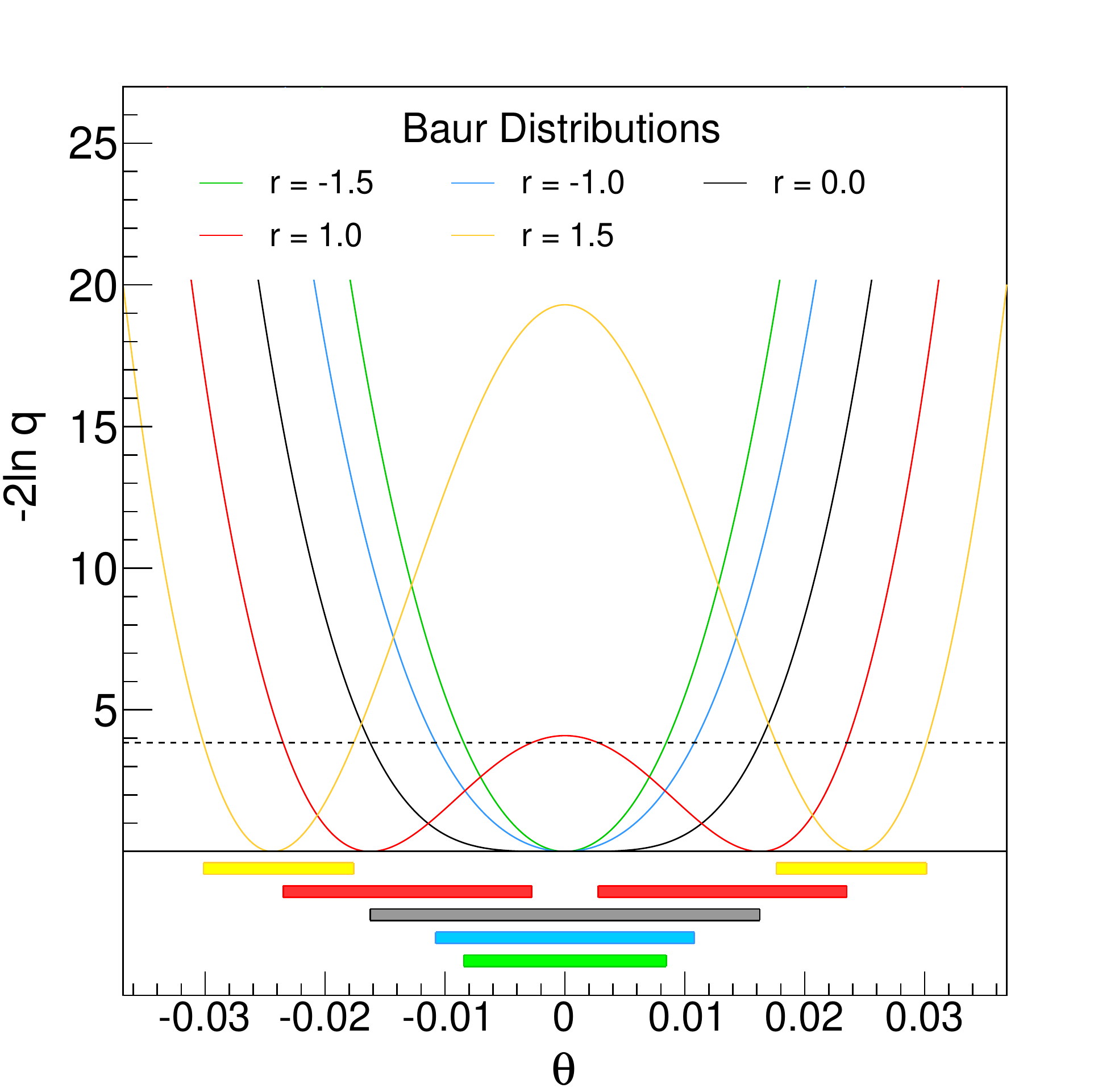}}
  \vspace{0.15cm}
  \caption{The curves show $-2\ln q$ for Baur distributions with $r~\in~\{0, \pm 1, \pm 1.5\}$ being used in turn as the observation. The boxes in the lower part indicate the corresponding confidence intervals at 95\% CL as determined by the delta likelihood ratio.}
  \label{fig:deltaLLR_baur} 
\end{figure}

When the linear term in the signal prediction is set to zero for all $x$, the Baur set has a very straight forward interpretation. For Baur distributions with $r\geq 0$, the event yield, $B(x;r)$, is greater than or equal to the SM expectation for all $x$, which means that these distributions are in the region described by the model. However, for Baur distributions with $r < 0$, the event yield is lower than the SM expectation for all $x$, and hence these Baur distributions are in the region which is not described by the model. 

When allowing a non-zero linear term in the signal prediction, the value of the lower bound on the signal prediction is lower than the SM expectation and thus the interpretation of the Baur set is different. The lower bound still persists, but it is shifted away from $\theta=0$ and is in general at different values of $\theta$ for different $x$. 

Figure~\ref{fig:baurset} shows Baur distributions for the binned observable for $r~\in~\{0, \pm 1, \pm 1.5\}$, with the value of $\sigma_\theta^{\rm ref}$ being determined by the confidence interval given in section \ref{sec:deltaLLR}.

\section{Confidence intervals for the Baur set}\label{sec:comparison}

This section compares the confidence intervals produced by the four different statistical methods introduced in section \ref{sec:methods} when a subset of Baur distributions are used in turn as the observation. The Baur distributions are made from the binned observable, and the linear term in the signal prediction is zero for all bins.

\begin{figure}
  \centering
  \vspace{-0.2cm}
  \resizebox{\columnwidth}{!}{\includegraphics{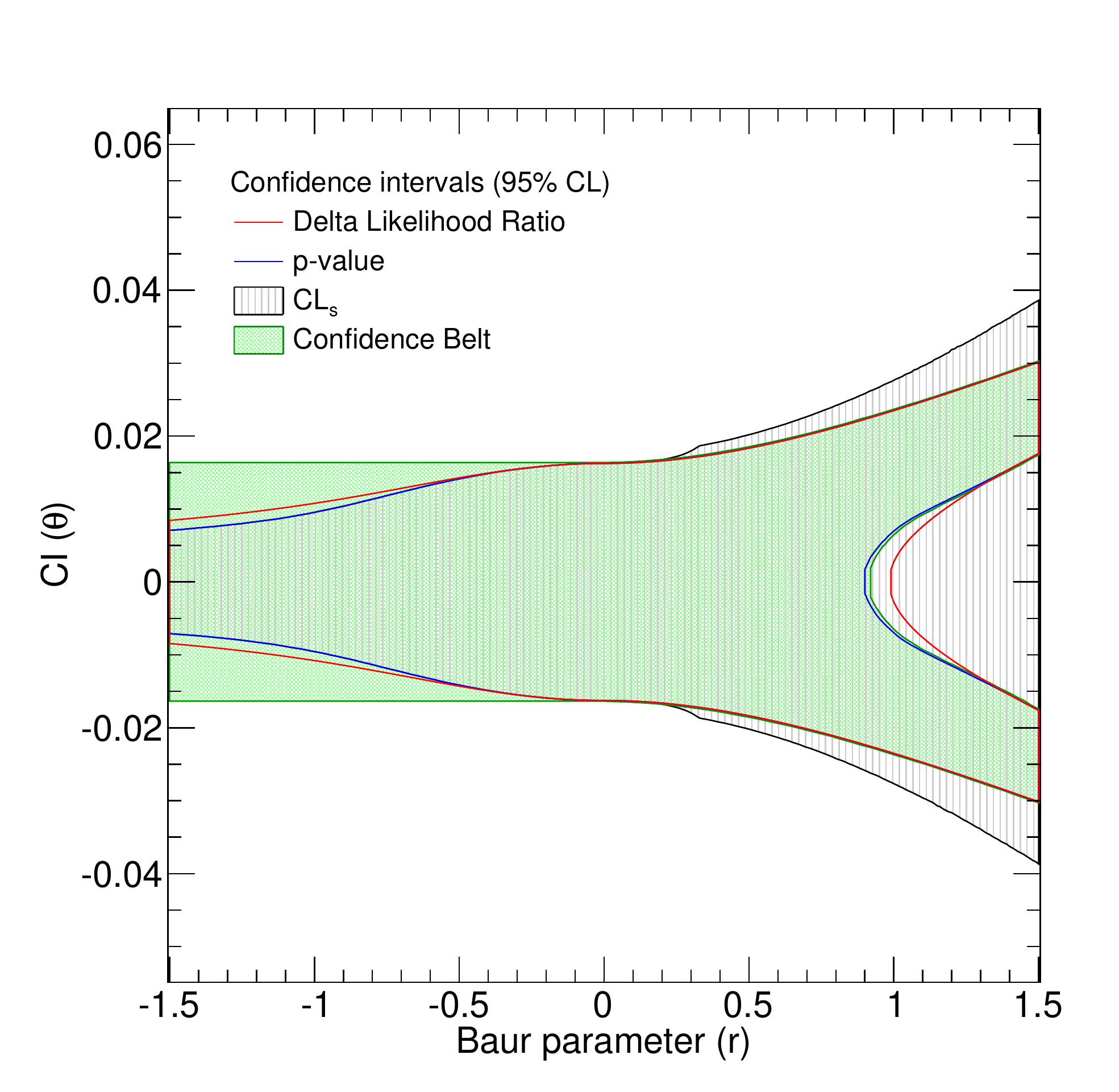}}
  \caption{Comparison of the four statistical methods for computing confidence intervals (CI). The confidence intervals at 95\% CL are shown as function of the Baur parameter, $r$, in the range $r=[-1.5,1.5]$.}
  \label{fig:comparison_baur}
\end{figure}

In order to illustrate the procedure, figure~\ref{fig:deltaLLR_baur} shows the confidence intervals as determined by the delta likelihood ratio when the Baur distributions with $r~\in~\{0, \pm 1, \pm 1.5\}$ are used in turn as the observation. The solid curves in the upper part in figure~\ref{fig:deltaLLR_baur} shows $-2\ln q$ for each of the Baur distributions. The intersections between the curves and the dashed horizontal line at 3.84 give the confidence intervals at 95\% CL which are shown in the lower part of figure \ref{fig:deltaLLR_baur} in corresponding colours. 

It is seen that the confidence intervals for the largest values of the Baur parameter ($r=1$ and $r=1.5$) consist of two disjoint intervals due to the corresponding $-2\ln q$ curves having two distinct minima. The two minima originate from the quadratic dependence on $\theta$ in the signal prediction. The reason they are symmetric around $\theta = 0$ and have equal depth is that the linear term in the signal prediction is zero for all bins. For $r<0$, there is only one minimum, $\hat{\theta}=0$. For these values of the Baur parameter, it is seen that the $-2\ln q$ curves become narrower as $r$ decreases, effectively decreasing the size of the confidence intervals. 
 
Similarly, confidence intervals as function of the Baur parameter can be computed for the other methods. The comparison between all methods is given in figure~\ref{fig:comparison_baur} which shows the confidence intervals when Baur distributions for values of $r$ in the range $r=[-1.5,1.5]$ are used in turn as the observation. A number of differences between the methods are clearly seen and these will serve as the basis for the discussion in the remainder of this section.

The first and main difference to be addressed is that the confidence intervals from the confidence belt remain constant for negative $r$ while the alternative methods give confidence intervals which are smaller as $r$ decreases. 

The intervals from the confidence belt remain constant because the maximum likelihood estimators are the same for all values of the Baur parameter below zero, namely $\hat{\theta}=0$, as also indicated in the upper part of figure \ref{fig:deltaLLR_baur}. Therefore it is the same intersection with the confidence belt, i.e. at $\hat{\theta} = 0$, which gives the confidence intervals for $r<0$. 

The alternative methods produce smaller intervals because $-2\ln q$ becomes narrower as $r$ decreases below zero as shown in the upper part of figure \ref{fig:deltaLLR_baur}. Evidently, this is also correlated with an increasing disagreement between the observation and the best fit. Since such a disagreement is described in terms of the goodness-of-fit, it indicates that the goodness-of-fit is encoded in the shape of the likelihood function for $r<0$.

In order to support this statement more quantitatively, the shape of $-2\ln q$ for Baur distributions with $r<0$ is examined by considering the simplified case where only the total number of events is used to estimate the parameter $\theta$, i.e. focusing on a single-bin observable. In this case, the likelihood is given by the Poisson probability of observing $n$ events with an expectation of $\mu(\theta)$, where $\mu(\theta)$ depends quadratically on $\theta$,
\begin{eqnarray}
  \mathcal{L} &=& \frac{\mu^n(\theta)}{n!}e^{-\mu(\theta)}.
\end{eqnarray}

In order to examine how the shape of $-2\ln q$ changes as function of $n$, or equivalently as function of $r$, the quantity $R(n;\theta)$ is defined, for a given $\theta$, as the difference in $-2\ln q$ for observing $n$ and $n_{\rm SM}$ events, respectively, 
\begin{eqnarray}
  R(n;\theta) &\equiv& [-2\ln q(n;\theta)] - [-2\ln q(n_{\rm SM};\theta)],
\end{eqnarray}
where $n_{\rm SM}$ refers to the expected number of events from the SM. The quantity $R(n;\theta)$ effectively describes how the shape of $-2\ln q$ varies for different observations, $n$.

Investigating the scenario where $n\leq n_{\rm SM}$, which corres-ponds to $r<0$, and using that the SM expectation is equal to the value of the lower bound on the signal prediction, i.e. $\hat{\theta}_n~=~\hat{\theta}_{n_{\rm SM}}~=~0$, it can be shown that $R(n;\theta)$ is given by
\begin{eqnarray}\label{Rn}
  R(n;\theta) &=& 2\left(n_{\rm SM}-n\right)\ln\left(\frac{\mu(\theta)}{n_{\rm SM}}\right) \quad , \quad n\leq n_{\rm SM}.
\end{eqnarray}

\begin{figure}
  \centering
  \vspace{-0.6cm}
  \resizebox{1.08\columnwidth}{!}{\includegraphics{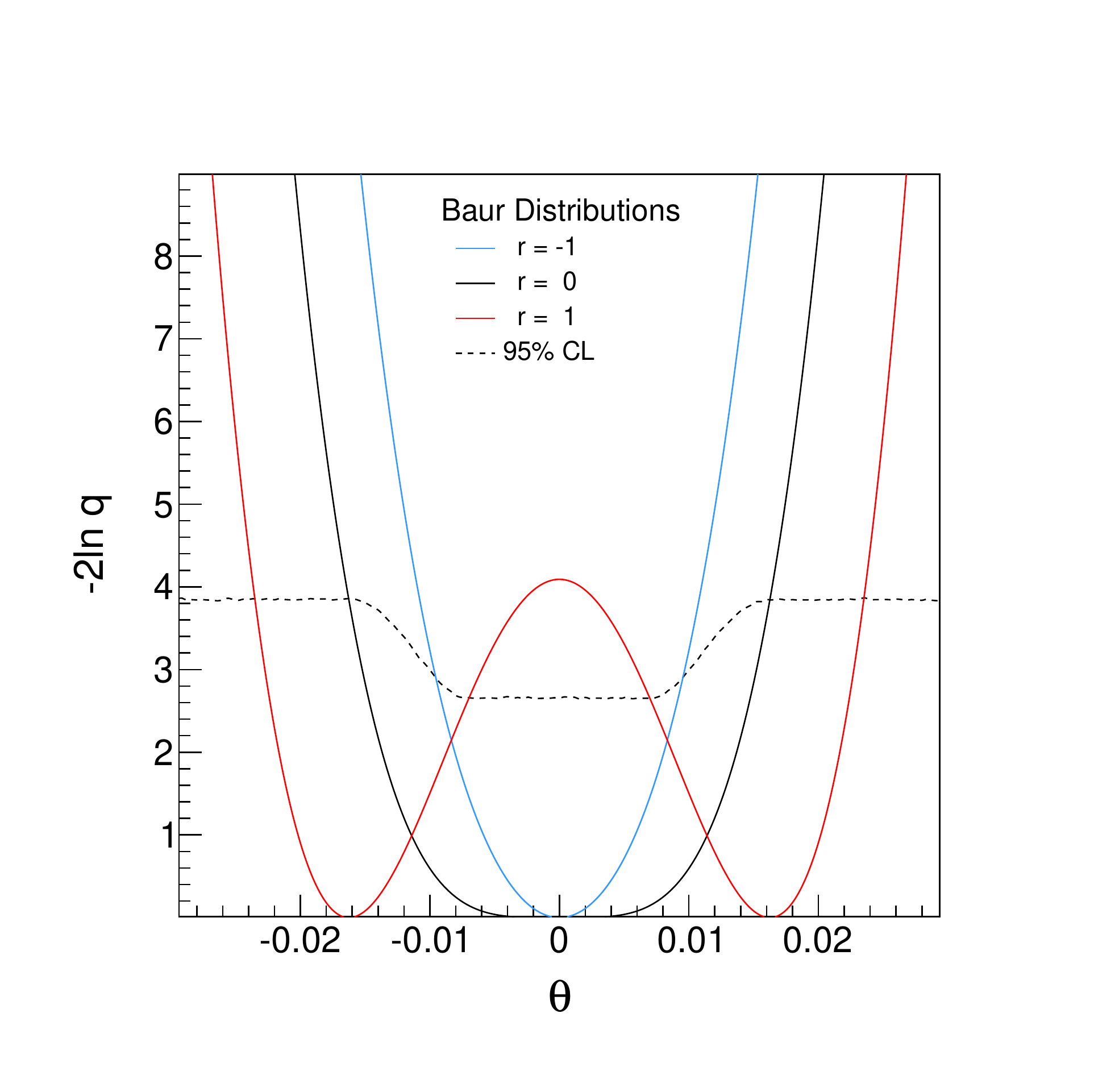}}
  \caption{The curves show $-2\ln q$ for observations given by Baur distributions with $r~\in~\{0,\pm 1\}$, respectively. The dashed line displays the 95\% CL contour line as determined by the pseudo experiments. The intersections between the curves and the contour line gives the confidence intervals for the $p$-value method.}
  \label{fig:pvalue_contour}
\end{figure}

Due to the quadratic dependence on $\theta$, $\mu(\theta)$ is greater than or equal to $n_{\rm SM}$ for all values of $\theta$, and consequently, $R(n;\theta)$ is positive and increasing linearly with decreasing $n$ for any $\theta~\neq~0$. This explains why the shape of $-2\ln q$ becomes narrower for decreasing $r$.

The corresponding goodness-of-fit as function of $n$ is described by the chi-square test statistic,
\begin{eqnarray}
  \chi^2(n) &=& \frac{(n - \mu(\hat{\theta}_n))^2}{\sigma^2} \ =\ \frac{(n - n_{\rm SM})^2}{n_{\rm SM}},
\end{eqnarray}
for $n\leq n_{SM}$. 

It seen that  $R^2(n;\theta)$ and the chi-square are directly proportional to each other,
\begin{eqnarray}
  R^2(n;\theta) \propto \chi^2(n),
\end{eqnarray}
which means that there is a direct link between the shape of the likelihood function and the goodness-of-fit for scenarios where fewer events are observed than what is predicted by the SM.

Consequently, any statistical method which relies on the shape of the likelihood function will encode the goodness-of-fit measure into the confidence interval which is clearly undesireable. Since the alternative methods for computing confidence intervals explicitly depend on the shape of the likelihood function, they will provide biased intervals which, as seen in figure \ref{fig:comparison_baur}, over-constrain the parameter when fewer events are observed than what is expected from the SM. 

\begin{figure}
  \centering
  \vspace{0.08cm}
  \resizebox{1.1\columnwidth}{!}{\includegraphics{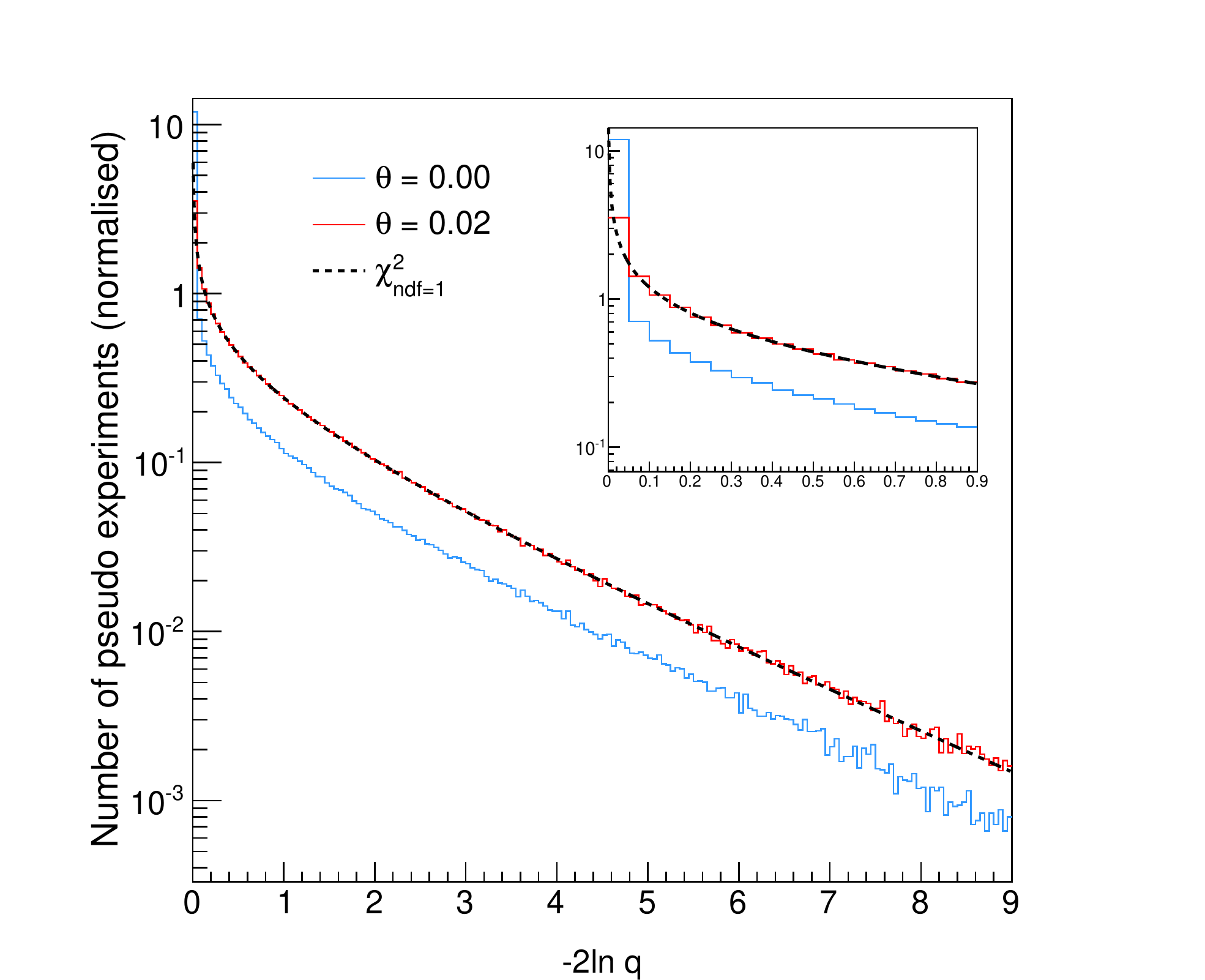}}
  \vspace{0.34cm}
  \caption{Distributions of $-2\ln q$ for two value of the parameter, $\theta=0$ (solid blue line) and $\theta=0.02$ (solid red line), and the distribution of the chi-square for one free parameter (dashed black line). The inset figure is a zoom-in on the lower region. }
  \label{fig:chi2}
\end{figure}

Another striking difference between the statistical methods displayed in figure \ref{fig:comparison_baur}, is that the $CL_s$ method gives considerably larger intervals than the other methods for large positive values of the Baur parameter, which notably also do not separate into two disjoint intervals. These features are due to the fact that $-2\ln q$ has a local maximum at $\theta=0$, the value of which increases with increasing $r$ (see the upper part of figure \ref{fig:deltaLLR_baur}). Consequently, the corresponding $p$-values for the SM, i.e. $CL_b$, decrease and the confidence intervals grow in size and, by construction, never split into two. The fact that the $CL_s$ method gives larger intervals for these values of $r$ is not surprising since the method by construction is meant to overestimate the intervals.

It is also interesting that for $r<0$ the confidence intervals produced by the $CL_s$ method are identical to those produced by the $p$-value method. Naively, one would expect the $CL_s$ method to expand the confidence intervals in situations where the SM expectation is disfavoured by the observation, as is the case for these Baur distributions. However, due to the lower bound on the signal prediction, the minimum of $-2\ln q$ is at $\theta=0$ and hence the $p$-value for the SM hypothesis for an observation given by a Baur distribution with $r<0$ is misleadingly equal to one, $CL_b=1$. As a result, the quantities denoted $p(\theta)$ and $CL_s$ in equations \ref{eqn:pvalue} and \ref{eqn:cls}, respectively, are the same and thus the $p$-value and $CL_s$ methods provide identical confidence intervals.

We now address two more subtle differences between the methods which are seen in figure \ref{fig:comparison_baur}. These will be explained in detail since it gives a good understanding of the basic mechanisms at play which are important for the overall description of the statistical methods.

The first is that the $p$-value and $CL_s$ methods provide smaller confidence intervals than the delta likelihood method for $r<0$. The second is that the $p$-value and delta likelihood methods disagree on the value of $r$ where the confidence interval breaks into two disjoint intervals. For the delta likelihood method this occurs by construction at $r=1$, while the $p$-value method also produces two disjoint intervals for values of $r$ slightly below one.

In order to examine these observations in more detail, figure~\ref{fig:pvalue_contour} shows $-2\ln q$ for observations given by three values of the Baur parameter, $r\in\{0, \pm 1\}$, (solid curves) superimposed on the 95\% CL contour line as determined by the pseudo experiments (dashed line), i.e. the line above which 5\% of the pseudo experiments fall for a given $\theta$. From this figure, the confidence intervals at 95\% CL for the $p$-value method are given by the intersections between the 95\% CL contour line and the curves showing $-2\ln q$.

It is seen that for large $|\theta|$, corresponding to the region far away from the bound, the 95\% CL contour line agrees with 3.84. However, for small $|\theta|$ the line has a shift towards a lower plateau due to the boundary. When investigating the distributions of $-2\ln q$ for two values of $\theta$ (see figure \ref{fig:chi2}), it is seen that the shift is due to many pseudo experiments having $-2\ln q = 0$. The distribution of the chi-square for one free parameter is superimposed (dashed curve) and it shows perfect agreement with the distribution of $-2\ln q$ for the large value of $|\theta|$ (the red histogram) as expected.

The  modification of the distribution of $-2\ln q$ in figure \ref{fig:chi2} (blue histogram), and the corresponding downward shift in the 95\% CL contour line in figure \ref{fig:pvalue_contour} (dashed line), occur since the pseudo experiments not described by the model have $\hat{\theta} = 0$. Consequently, when scanning through values of $\theta$ close zero, the value of $-2\ln q(\theta)$ for these pseudo experiments is also close zero, and for $\theta = 0$ it is identically zero. The fraction of these pseudo experiments grows as $\theta$ approaches zero at which point it reaches approximately one half. The lower plateau manifests itself when all of these pseudo experiments have migrated below the 95\% CL contour line. For larger values of $|\theta|$, where the pseudo experiments only rarely probe the region not described by the model, the value of $-2\ln q$ is not significantly affected and thus the 95\% CL contour line agrees with 3.84.

The downward shift in the 95\% CL contour line in figure \ref{fig:pvalue_contour} means that the intersections between this line and the curves showing $-2\ln q$ occur at different values of $\theta$ than the corresponding intersections between these curves and a line at 3.84. Consequently, the delta likelihood ratio method provides larger intervals for Baur distributions with $r<0$ than the $p$-value and $CL_s$ methods, and the $p$-value method produces two disjoint intervals for slightly smaller values of $r$ compared to the delta likelihood method.

As a final remark, it should be noted that while the Baur distributions efficiently illustrate a number of differences between the statistical methods, the situation is, in general, more complicated since the data does not necessarily have the same trend for all values of the observable. For example, a deficit of events with respect to the SM expectation in a region less sensitive to the parameter can be compensated by a surplus in a more sensitive region. This aspect complicates the situation considerably, and in fact there is no way to know if a confidence interval computed with one of the alternative methods is biased or not without explicitly also computing it with the confidence belt. This is particularly interesting as it differs from the situation where the signal prediction only depends linearly on the parameter of interest. In this case, the confidence belt corresponds to the acceptance region of the hypothesis test and thus the $p$-value method will always give the same result as the confidence belt. As demonstrated here, this is not the case when the signal prediction depends quadratically on the parameter of interest.

\section{Non-zero interference}
\label{sec:interference}

In the previous sections, it was assumed that the linear term in equation \ref{eqn:signal} was absent. This section will address the general case with a non-zero linear term. 

For the case of effective field theories, the linear term $a_1(x)$ corresponds to a interference term and can be written as
\begin{eqnarray}\label{eqn:a1}
  a_1(x) &=& 2\sqrt{a_0(x)a_2(x)}\cos(\Delta\phi(x)),
\end{eqnarray}
where $\Delta\phi(x)$ is the phase difference between the amplitudes $A_\mathrm{SM}(x)$ and $A_\mathrm{BSM}(x)$. 

The unknown dependence of $a_1(x)$ is described entirely by the phase difference through $\cos(\Delta\phi(x))$. Since cosine is limited to the range $[-1,1]$, the size of $a_1(x)$ is less than or equal\footnote{Note that for the extreme case of $\cos(\Delta\phi) = \pm 1$, corresponding to a minimum signal prediction of exact zero, the Poisson likelihood is not defined.} to $2\sqrt{a_0(x)a_2(x)}$.

In order to test the effects of a non-zero linear term, the signal prediction is modified using equation \ref{eqn:a1}, dropping the $x$ dependence in $\cos(\Delta\phi)$ for simplicity. A range of values for $\cos(\Delta\phi)$ is considered.

\begin{figure}
  \centering
  \resizebox{\columnwidth}{!}{\includegraphics{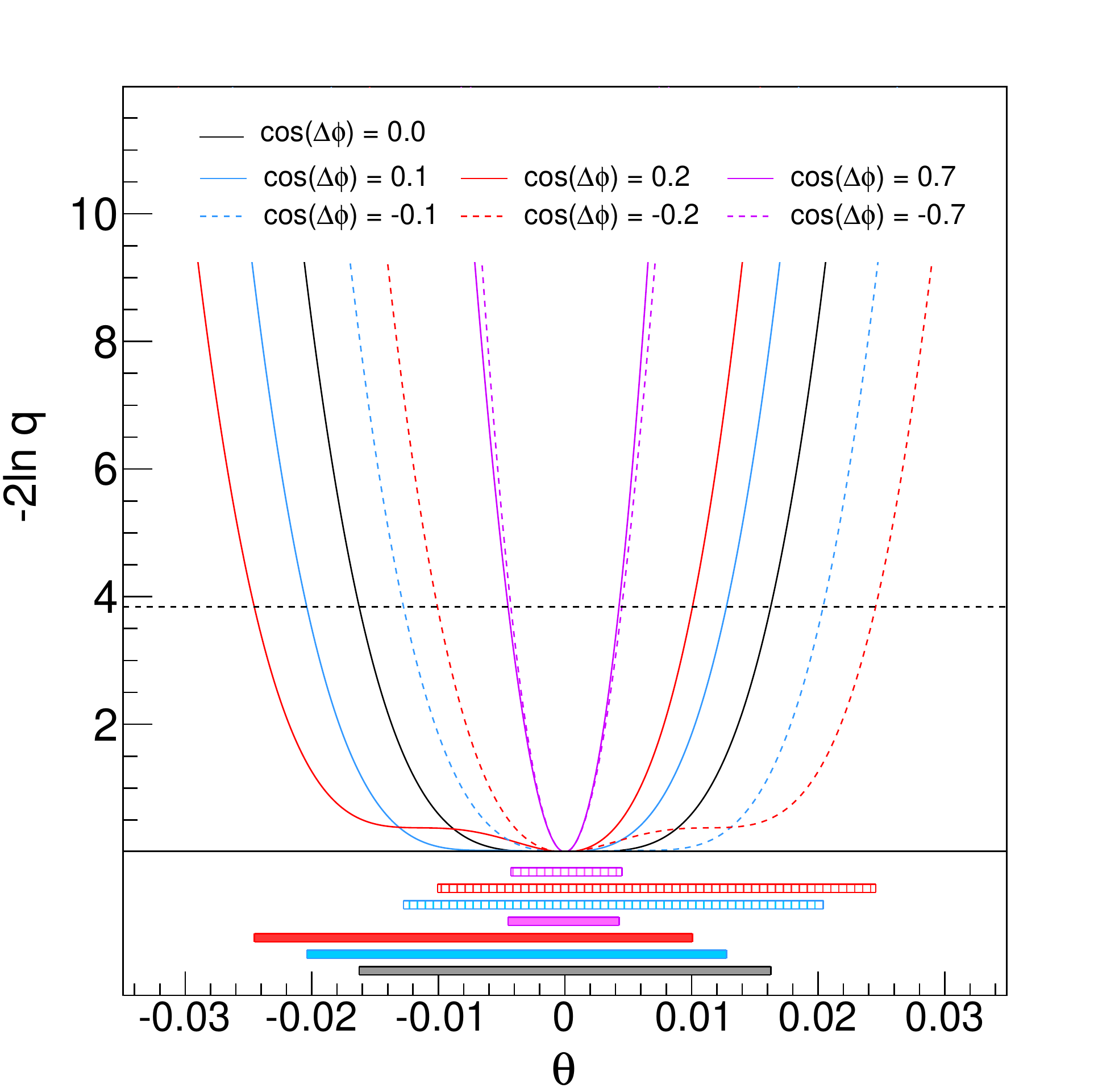}}
  \caption{The curves show $-2\ln q$ for different values of $\cos(\Delta\phi)$ in the signal prediction, and with the SM expectation used as the observation in all cases. The boxes in the lower part indicate the corresponding confidence intervals at 95\% CL as determined by the delta likelihood ratio.}
  \label{fig:deltaLLR_linTerm} 
\end{figure}

\begin{figure*}[t!]
  \centering
  \vspace{-0.3cm}
  \subfloat[][$\cos(\Delta\phi)=0.1$.]{\resizebox{0.855\columnwidth}{!}{\includegraphics{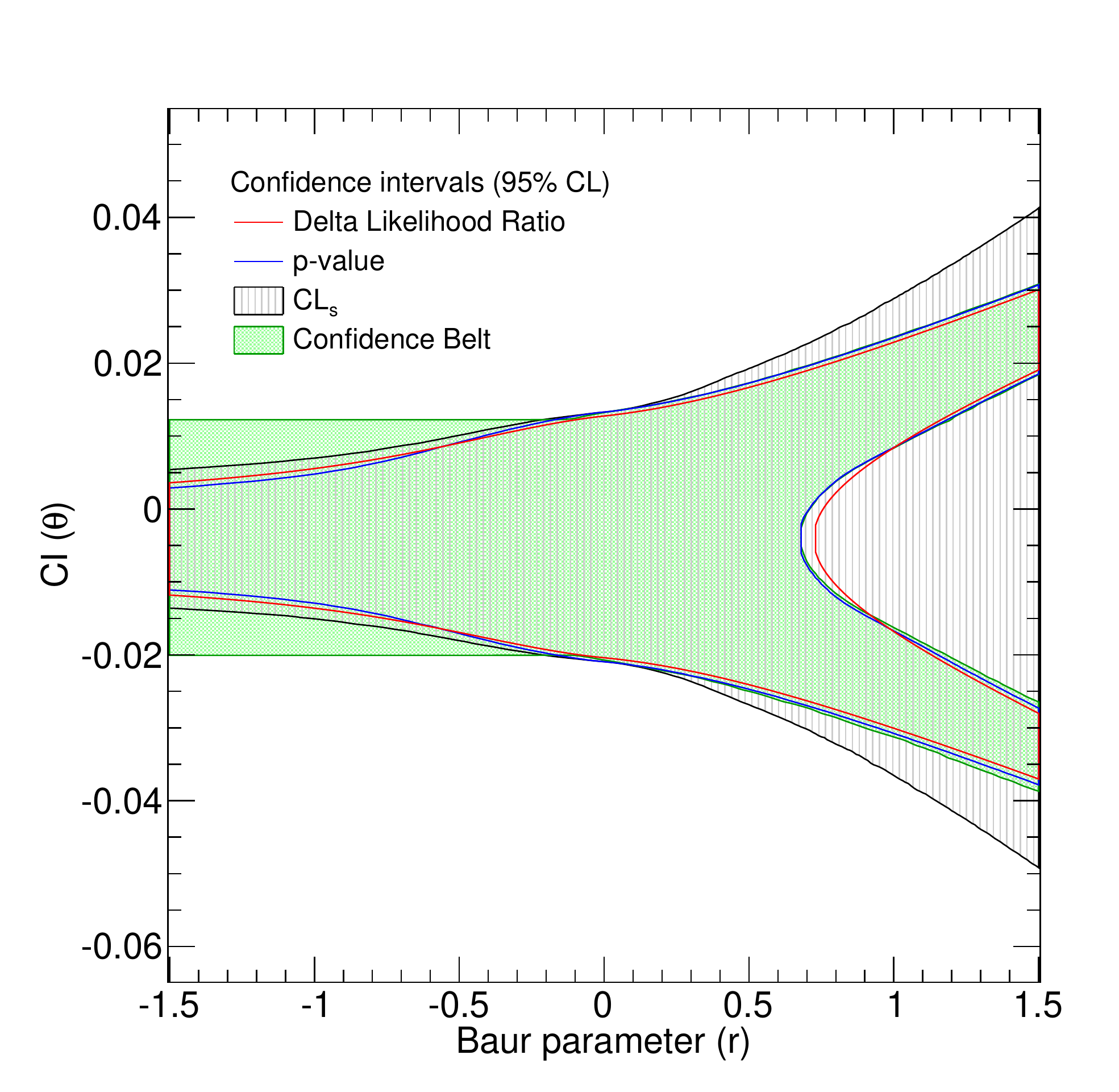}}\label{fig:comparison_baur_cosPhi_0.1}}
  \hspace{0.5cm}  
  \subfloat[][$\cos(\Delta\phi)=-0.1$.]{\resizebox{0.855\columnwidth}{!}{\includegraphics{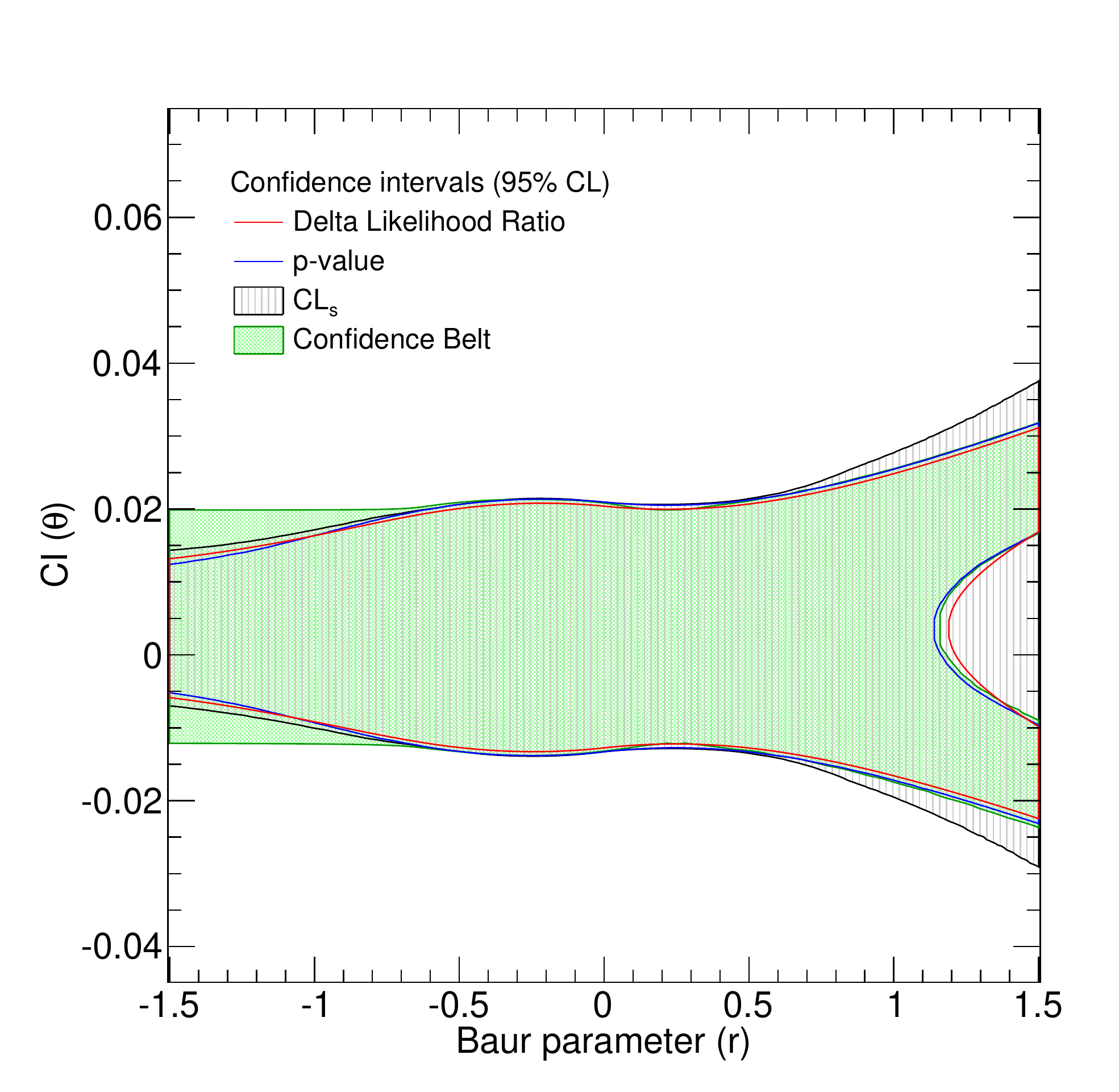}}\label{fig:comparison_baur_cosPhi_-0.1}}
  \\
  \vspace{-0.3cm}
  \subfloat[][$\cos(\Delta\phi)=0.2$.]{\resizebox{0.855\columnwidth}{!}{\includegraphics{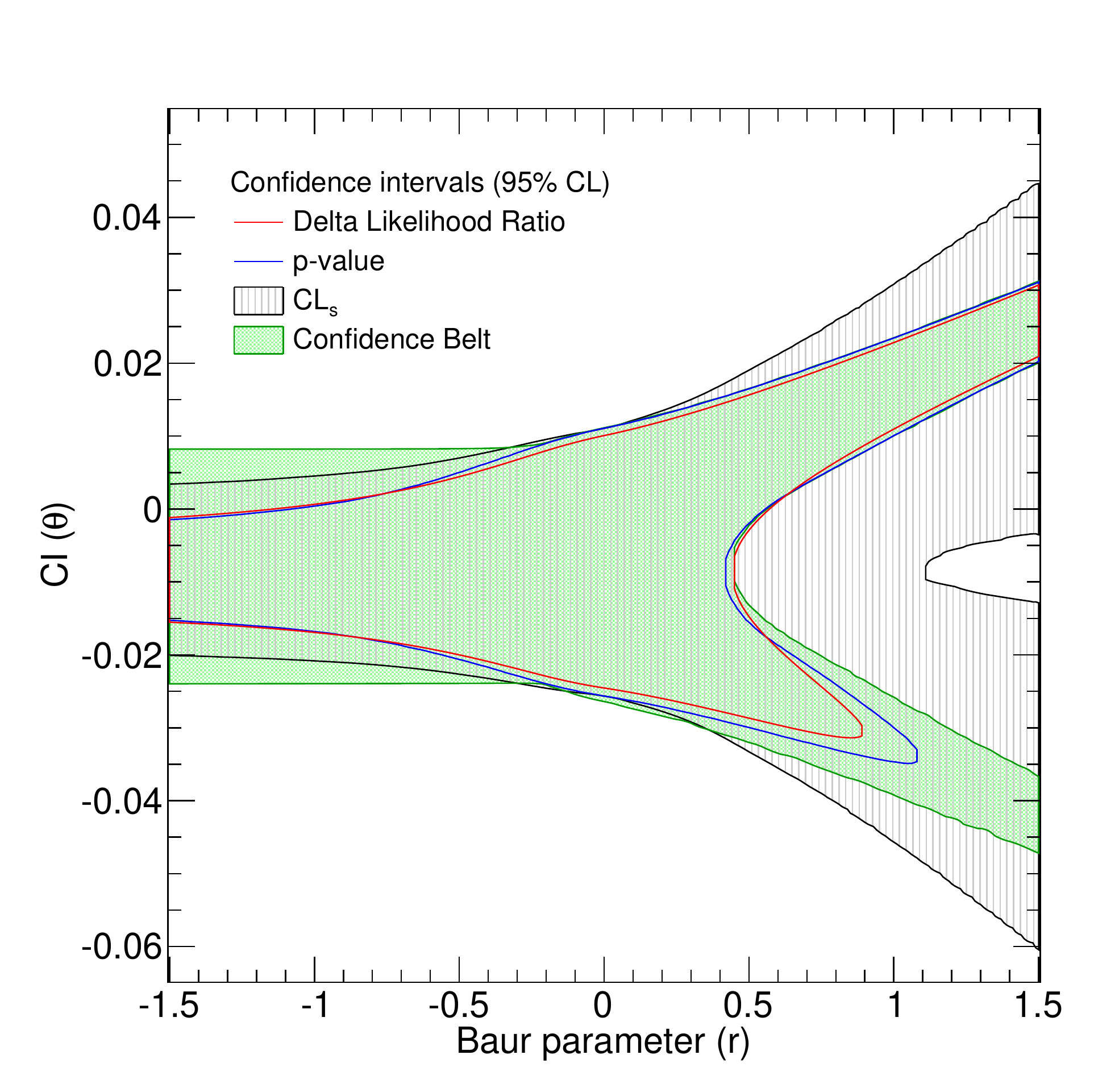}}\label{fig:comparison_baur_cosPhi_0.2}}
  \hspace{0.5cm}
  \subfloat[][$\cos(\Delta\phi)=-0.2$.]{\resizebox{0.855\columnwidth}{!}{\includegraphics{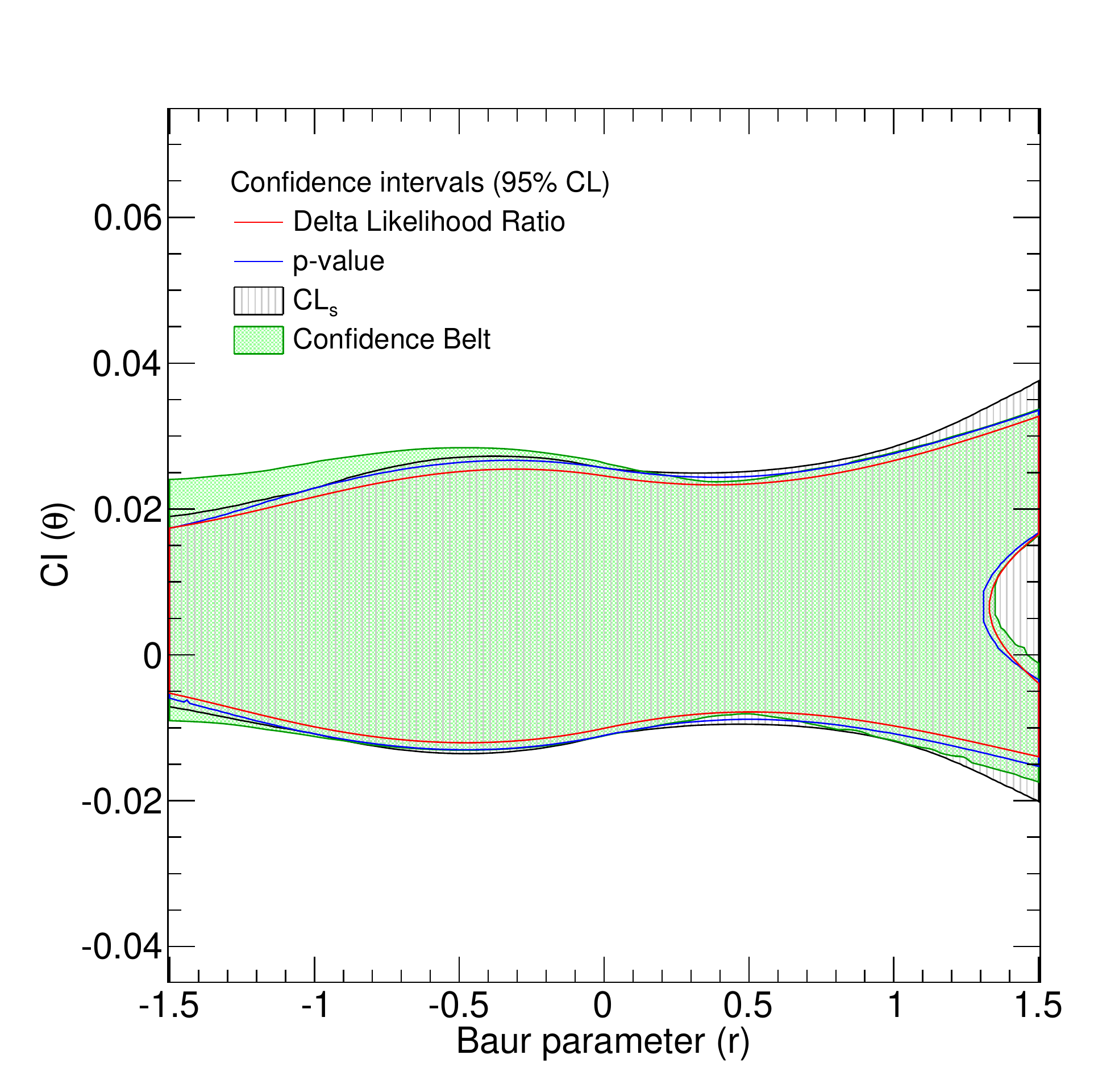}}\label{fig:comparison_baur_cosPhi_-0.2}}
  \caption{Confidence intervals at 95\% CL as function of $r$ in the range $r=[-1.5,1.5]$, using four different values for $\cos(\Delta\phi)$ in the signal prediction.}
  \label{fig:comparison_interference}
\end{figure*}

In order to give an idea of how the likelihood function is affected when the model is modified, figure \ref{fig:deltaLLR_linTerm} shows the curves for $-2\ln q$ for an observation at the SM expectation using seven different values of $\cos(\Delta\phi)$ in the signal prediction, $\cos(\Delta\phi)~\in~\{0,\pm 0.1, \pm 0.2, \pm 0.7\}$, corresponding to 0\%, 10\%, 20\% and 70\% of the maximal interference. It is seen that as the size of the negative (positive) interference terms increase, there is a shift towards positive (negative) values of $\theta$ in the $-2\ln q$ curves and that a shoulder appears on the right (left) hand side of the minimum. The intersections between the curves and the dashed horizontal line at 3.84 give the confidence intervals at 95\% CL as determined by the delta likelihood method which are shown in the lower part of figure \ref{fig:deltaLLR_linTerm} in corresponding colours. As the shoulder moves above the line at 3.84, which is the case for the extreme value $\cos(\Delta\phi) = \pm 0.7$, the confidence intervals get smaller and becomes increasingly symmetric around $\theta=0$. This reflects the fact that the linear term in the signal prediction begins to dominate. 

The comparison between the statistical methods for different observations are done using Baur sets constructed for four for different values of $\cos(\Delta\phi)$ in the signal prediction, $\cos(\Delta\phi) \in \{\pm 0.1, \pm 0.2\}$. Figures \ref{fig:comparison_baur_cosPhi_0.1}-\ref{fig:comparison_baur_cosPhi_-0.2} show the confidence intervals when the observation is given by the Baur distributions for values of $r$ in the range $r~=~[-1.5,1.5]$ for each of the four Baur sets, respectively. 

For all four Baur sets, clear trends for the statistical methods are observed. First, it should be mentioned that the qualitative differences between the graphs for positive versus negative $\cos(\Delta\phi)$ are due to the specific choice of the sign of $\sigma_{\rm ref}$ in equation \ref{eqn:sigma_ref}. If changing the sign of $\sigma_{\rm ref}$, the features are reversed between positive and negative $\cos(\Delta\phi)$. For instance, it is seen that for positive values of $\cos(\Delta\phi)$ the sizes of the confidence intervals are strictly increasing with $r$ (until the point where they break into two disjoint intervals), whereas for negative values of $\cos(\Delta\phi)$ there is an intermediate range in $r$ around $r=0$ where the sizes of the confidence intervals descrease with $r$. This is directly related to the sign of $\sigma_{\rm ref}$ and the effect would be reversed if the sign was changed. 

Addressing the differences between the methods, it is seen in figure \ref{fig:comparison_baur_cosPhi_0.2} that the otherwise defining feature of having two disjoint confidence intervals for large values of the Baur parameter does not apply to the delta likelihood and the $p$-value methods. The reason is that for a combination of sufficiently large $\cos(\Delta\phi)$ and $r$, there is sensitivity to the sign of the parameter. More specifically, it means that the two minima in $-2\ln q$ for the Baur distributions at large $r$ are separated to an extend which makes the non-global minimum lie above the threshold for a 95\% CL. Thus, only one confidence interval is produced, and this will always be the upper one in the figure due to the way the Baur distributions are defined. Hints of this trend can also be seen in figures \ref{fig:comparison_baur_cosPhi_0.1}, \ref{fig:comparison_baur_cosPhi_-0.1} and \ref{fig:comparison_baur_cosPhi_-0.2} where the lower intervals produced by delta likelihood and $p$-value methods are slightly smaller than the corresponding intervals given by the confidence belt for large values of $r$. As is seen in figure \ref{fig:comparison_baur_cosPhi_0.2}, the two methods do not agree exactly on where the transition region for producing one or two intervals is, only that it is around $r=1$. This arise since the distribution of $-2\ln q$ does not exactly follow a chi-square distribution for all $\theta$. 

In contrast, it is seen that the confidence belt method for all four values of $\cos(\Delta\phi)$ produces two disjoint confidence intervals for large $r$. The reason is that the cross-like shape of the confidence belt persists for all four values of $\cos(\Delta\phi)$. However, it should be mentioned that the density of pseudo experiments is different in the two diagonal branches in the confidence belt when $\cos(\Delta\phi) \neq 0$. The branch with the negative slope in figure \ref{fig:neyman_b} has a much lower fraction of the pseudo experiments, the trend being that the density decreases with increasing $|\cos(\Delta\phi)|$. In fact, for high enough values of $|\cos(\Delta\phi)|$, the branch with negative slope in the confidence belt will disappear, at which point the confidence belt only produces one confidence interval. 

The discrepancy between the delta likelihood, $p$-value and the confidence belt methods is interesting since it implies that the former two do not manage to fully map the relation between the parameter of interest and its maximum likelihood estimator. This is best understood by considering the level of information used by the $p$-value method when a pseudo experiment is performed for a given $\theta$. As explained in section \ref{sec:pvalue}, the $p$-value method counts the number of pseudo experiments where the value of $-2\ln q$ is larger than it is for the observation. However, only using the value of $-2\ln q$ does not encapsulate the fact that there are potentially two minima in $-2\ln q$ for each pseudo experiment, and that the global minimum fluctuates between the two from one pseudo experiment to the next. Consequently, the $p$-value and delta likelihood methods over-constrain the parameter.

Another interesting feature is that for negative values of the Baur parameter, the $CL_s$ method expands the confidence intervals compared to the $p$-value method. This effect becomes more distinct as $|\cos(\Delta\phi)|$ increases. The reason is that the minimum in the signal prediction is not at the SM value, $\theta = 0$, but rather shifted towards positive (negative) values of $\theta$ for negative (positive) values of $\cos(\Delta\phi)$. Consequently, the $p$-value for the SM, $CL_b$, is less than one and the confidence interval gets expandend compared to the interval from the $p$-value method. 

Finally, as seen in figure \ref{fig:comparison_baur_cosPhi_0.2}, the $CL_s$ method for \linebreak $\cos(\Delta\phi)=0.2$ produces two separated intervals for large values of $r$. In order to understand this, it should first be recalled that $-2\ln q$ has a local maximum for large values $r$ (see e.g. figure \ref{fig:deltaLLR_baur}, red and yellow graphs). For $|\cos(\Delta\phi)|$ above a certain value, the difference between the $p$-values at the SM, $CL_b$, and at the local maximum, $CL_{s+b}(\theta_{\rm max})$, becomes so large that a region around $\theta_{\rm max}$ is not included in the confidence interval, i.e. $CL_s(\theta_{\rm max}) < 1 - \alpha$, where $\alpha$ denotes the confidence level. Consequently, this  gives two disjoint confidence intervals on each side of $\theta_{\rm max}$.

\section{Conclusion}
\label{sec:conclusion}
The effective Lagrangians approach used in most model independent searches for BSM physics introduces a bound on the signal prediction due to a quadratic parameter dependence in the differential cross section. The bound is typically a lower bound due to the non-renormalisability of the BSM terms and is often located close to or at the SM expectation for physics cases such as anomalous Higgs couplings, anomalous trilinear or quartic gauge couplings.

While the original frequentist approach for determining confidence intervals, known as the confidence belt, explictly computes the mapping of the observation in data into a subset of values for the true parameter, thus giving the correct frequentist coverage for all observational scenarios, it is demonstrated that statistical methods currently employed at the LHC, i.e. the delta likelihood, the $p$-value and the $CL_s$ methods, systematically over-constrain the parameter when data shows distinct fluctuations into the region which is not described by the model. 

The presence of a interference term between the SM and the BSM amplitudes improves the ability of the model to describe data in the vicinity of the SM. However, it also shows that the delta likelihood, the $p$-value and the $CL_s$ methods in general fail to map the observation in data into the full subset of values for the parameter, even for observations which are fully described by the model. Consequently, the experimental sensitivity to interference terms depends on statistical procedures.

It should be emphasized that the present findings show that the usual correspondance between the confidence belt and the hypothesis test performed in the $p$-value method, i.e. that the former constitutes the acceptance region of the latter, is not true for the case where the parameter of interest enters quadratically in the signal prediction. In fact, this statement is true for any functional dependency on the parameter which introduces a region not described by the model. For physics scenarios where this is the case, the delta likelihood, the $p$-value and the $CL_s$ methods are not guaranteed to provide the correct frequentist coverage.

\subsection*{Acknowledgements} 
The authors are grateful to Professor Emeritus J. D. Hansen for useful discussions and valuable suggestions related to this work.



\end{document}